\documentclass[reprint,aps,prd,nofootinbib]{revtex4-2}

\usepackage{amsmath,amssymb,epsfig,bm,amsfonts,yfonts,bbm,mathrsfs}
\usepackage{tensor,todonotes}

\begin{document}

\title{Conserved and non-conserved Noether currents from the quantum effective action}

\author{Stefan~Floerchinger}
\email{stefan.floerchinger@uni-jena.de}

\affiliation{
Theoretisch-Physikalisches Institut, Friedrich-Schiller-Universität Jena, Max-Wien-Platz 1, 07743 Jena, Germany
}

\affiliation{
Institut f\"{u}r Theoretische Physik, Ruprecht-Karls-Universit\"{a}t Heidelberg, 
Philosophenweg~16, 69120 Heidelberg, Germany
}

\author{Eduardo Grossi}
\email{eduardo.grossi@stonybrook.edu}
\affiliation{Center for Nuclear Theory, Department of Physics and Astronomy,
Stony Brook University, Stony Brook, New York 11794-3800, USA}

\begin{abstract}
The quantum effective action yields equations of motion and correlation functions including all quantum corrections. We discuss here how it encodes also Noether currents at the full quantum level. Interestingly, the construction can be generalized beyond the standard symmetry transformations that leave the action invariant. We also discuss an extended set of transformations, which change the action by a term that is locally known on the level of the quantum effective action. Associated to such extended gauge transformations are currents for which we obtain a divergence-type equation of motion, but they are not conserved. We call them non-conserved Noether currents. We discuss in particular symmetries and extended transformations associated to space-time geometry for relativistic quantum field theories. These encompass local dilatations or Weyl gauge transformation, local Lorentz transformations and local shear transformations. Together they constitute the symmetry group of the frame bundle GL$(d)$. The corresponding non-conserved Noether currents are the dilatation or Weyl current, the spin current and the shear current. In particular for the latter we obtain a new divergence-type equations of motion.
\end{abstract}

\maketitle

\section{Introduction}
The relation between the microscopic formulation of a quantum field theory, and the macroscopic formulation which includes the effect of quantum and statistical fluctuations, can be nicely discussed in terms of actions. The microscopic action $S[\chi]$ defines a theory at a microscopic scale or at very high momenta where quantum fluctuations are suppressed. This is the object that enters the functional integral. In classical situations where quantum fluctuations are negligible, the microscopic action yields directly the classical Euler-Lagrange equations and for this reason it is sometimes called classical action. The microscopic action depends of course on the field values but is otherwise independent of the state. Initial conditions enter only as boundary conditions for solutions to the equations of motion. 

In contrast to this, the one-particle irreducible or quantum effective action (see e.\ g.\ \cite{ZinnJustin:2002ru, Weinberg:1996kr, Piguet:1995er}, we recall its construction below) depends on field expectation values and can be used to derive various quantities for which all quantum corrections have already been taken into account. For example, the propagators and vertices obtained from functional derivatives of the one-particle irreducible effective action around a vacuum solution yield the full correlation functions and S-matrix elements when used in tree diagrams. 

Similarly, from the variation of the quantum effective action with respect to the fields one obtains renormalized equations of motion. It becomes already clear from these statements that the determination of the quantum effective action itself is typically a formidable task. In particular, it differs from the microscopic action through both perturbative and non-perturbative quantum and statistical corrections. There are several methods to take these corrections into account; one of them is the functional renormalization group \cite{Wilson:1971bg, Wegner:1972ih, Polchinski:1983gv, Wetterich:1992yh}.

One should remark here that the quantum effective action can, and will, contain terms of higher derivative order in the fields, and possibly non-local terms. All terms allowed by symmetries can appear and are typically present. Scaling arguments based on the classification of operators into relevant, irrelevant and marginal can sometimes be used in the vicinity of renormalization group fixed points, but that is not the generic situation. One should also remark that the quantum effective action depends in general on the quantum state through the boundary conditions of the functional integral. 

With the present paper we have two major goals. First, we want to clarify how expectation values and correlation functions of standard Noether currents \cite{Noether:1918zz}, associated to a continuous symmetry of a theory, can be obtained from the quantum effective action. This goal will be reached by using external gauge fields that are used to render the symmetry transformation local, and by then using these external gauge field to take functional derivatives to obtain expectation values and correlation functions of the currents.

The second goal is a generalization of the entire construction to an extended class of transformations, which have been called extended symmetries \cite{Canet:2014cta, Tarpin:2017uzn, Tarpin:2018yvs}. Indeed, it is possible to study continuous transformations under which the action of a theory is actually not invariant, and to still derive very useful relations, under certain circumstances. The class of transformations we have in mind here is such that the quantum effective action changes by a term that is locally known, because it is proportional to a field expectation value, or to the expectation value of a composite operator, to which an external source field has been coupled. Associated to such extended transformations are currents with a divergence-type equation of motion, which are, however, not conserved. Because their construction is still very close to the construction of Noether currents, we call them non-conserved Noether currents.

Let us emphasize why we believe it is useful to employ the quantum effective action to discuss conserved and non-conserved Noether currents. The quantum effective action differs from the microscopic action through the effect of quantum and statistical fluctuations, and these can make a crucial difference for the obtained currents. For example, in the presence of quantum anomalies it is possible that a current is conserved at the classical level, but not conserved when quantum fluctuations have been taken into account. It is even possible that a current seems to vanish when one attempts to derive it from the microscopic action, but that it is in fact non-vanishing when obtained from the quantum effective action. It has then contributions from quantum and statistical fluctuations only. We will present a concrete example where this happens in a scalar field theory below.

As a further motivation for our study one may think about fluid dynamics. Experience shows that in the macroscopic regime, i.\ e.\ for large time intervals and long distances, one can approximate quantum field dynamics often rather well by a variant of fluid dynamics \cite{Teaney:2009qa, Schafer:2009dj,CasalderreySolana:2011us}. This follows the rational that those degrees of freedom are important over long time intervals that are preserved from relaxation by conservation laws \cite{KADANOFF1963419, Hohenberg:1977ym, Floerchinger:2016gtl}. In practise, for a relativistic fluid, it is for reasons of causality not possible to take only the strictly hydrodynamical degrees of freedom (which are directly governed by conservation laws, e.\ g.\ energy- and momentum densities) into account, but some non-hydrodynamical fields must be propagated, as well \cite{Hiscock:1983zz,Hiscock:1985zz,Muronga:2003ta,Denicol:2008ha,Floerchinger:2017cii}. For example, in Israel-Stewart theory \cite{Israel:1979wp, Muller:1967zza}, for a fluid with conserved energy-momentum tensor but no additional conserved quantum number, these are the shear stress and bulk viscous pressure. Equations of motion for the latter are typically postulated in a phenomenological way. We will argue that these identities should follow from the additional non-conservation laws for the dilatation current and shear current. In some regards, the equations we find are close to those proposed in ref.\ \cite{Geroch:1990bw}.

An extension of fluid dynamic equations to include the spin tensor has been proposed and discussed in several recent publications \cite{Becattini:2011ev, Florkowski:2017ruc,Montenegro:2018bcf,Becattini:2018duy,Speranza:2020ilk,Garbiso:2020puw,Gallegos:2021bzp}, partly with the motivation to explain  the measurement of the polarization of the hyperon $\Lambda$ produced in high energy nuclear collisions \cite{STAR:2017ckg} (see ref.\ \cite{Becattini:2020ngo} for a review). 

Correlation functions of conserved and non-conserved currents are also of high interest. For example, they can be used to describe critical behaviour in the vicinity of phase transitions. In the context of linear or weakly non-linear response theory around equilibrium states they can be used to obtain transport properties, e.\ g.\ through the Kubo relations \cite{Kubo:1957mj, Mori:1959zz, Hosoya:1983id,Teaney:2006nc, Baier:2007ix, Moore:2010bu, Hong:2010at,Floerchinger:2011sc,Jeon:2015dfa}. 

We will discuss different examples for normal and extended symmetries related to the geometry of space-time for a relativistic quantum field theory. These transformations encompass local changes of coordinates (diffeomorphisms), but also transformations in the frame and spin bundles such as local Lorentz transformations, dilatations, and shear transformations. While diffeomorphisms can be seen as a conventional gauge transformation, which leaves the action invariant, the situation is more complicated for the other transformations. Local dilatations only leave the action invariant in a scale invariant theory (for which a conformal theory is an example), but more generally they should be seen as extended symmetry transformations. For local shear transformations and also local Lorentz transformations this is always the case, at least for the class of quantum field theories usually considered.

The currents we discuss encompass two versions of the energy-momentum tensor (a generalization of the canonical energy-momentum tensor to the quantum effective action and the symmetric energy-momentum tensor), the spin current, the dilatation or Weyl current, and the shear current. The latter three form together a tensor of rank three known as the hypermomentum tensor \cite{1976ZNatA..31..111H,1976ZNatA..31..524H,Hehl:1976kv,PhysRevD.17.428,Hehl:1994ue}.

Hypermomentum was initially investigated mainly in the context of modified theories of gravity, because it constitutes a natural source for non-Riemannian geometrical structures such as torsion and non-metricity \cite{Kibble:1961ba,Sciama:1964wt,1976ZNatA..31..111H, 1976ZNatA..31..524H, Hehl:1976kv, PhysRevD.17.428, Hehl:1994ue, Capozziello:1996bi, Puetzfeld:2004yg, Vitagliano:2010sr, Blagojevic:2012bc,BeltranJimenez:2018vdo,Shimada:2018lnm, Jimenez:2019ghw, Iosifidis:2020gth}. We will essentially obtain the currents of hypermomentum from varying these geometrical structures and observing the response of the quantum effective action.
However, we want to argue that hypermomentum is not only of interest in the context of somewhat speculative extensions of Einsteins theory of general relativity. In fact we believe that the spin current, dilatation current and shear current can all be non-zero for generic interacting quantum field theories in non-equilibrium situations. Even if these currents are apparently absent at the classical level, they can arise from quantum fluctuations, similar to quantum anomalies. We will also discuss an effective action for a scalar field with non-minimal coupling to gravity to underline this point.

This paper is organized as follows. In section \ref{sec:SymmetriesAndExtendedSymmetries} we recall the construction of the quantum effective action and discuss continuous symmetry transformations of quantum field theories. In particular we review the derivation of the Slavnov-Taylor identity as a constraint to the quantum effective action and point out how it generalizes in a useful way to a class of extended symmetry transformations beyond the normal symmetry transformations that leave the action invariant. In section \ref{sec:ConservedAndNon-conservedNoetherCurrents} we discuss how one can obtain Noether currents associated to continuous symmetries from the quantum effective action through the use of external gauge fields. Here we also point out how the construction can be generalized from the well known standard continuous symmetries to extended symmetry transformations for which we obtain divergence-type equations of motion for currents that are, however, not conserved. In section \ref{sec:Space-timeSymmetriesAndExtendedSymmetries} we concentrate then on space-time transformations. We first recall the standard discussion of general coordinate transformations or diffeomorphisms with the metric-compatible and torsion-free Levi-Civita connection. In a subsequent step we introduce a more general affine connection, which has also torsion and non-metricity, and we study on that basis an extended class of space-time transformations. The latter encompasses local Lorentz transformations for which the associated current is the well known spin current and Weyl gauge transformations with associated dilatation current, but also less known local shear transformations and their associated shear current. In section \ref{sec:NonMinimallyCoupledScalarField} we discuss as a concrete example a scalar quantum field theory. We illustrate there that the components of the hypermomentum current are non-vanishing as a consequence of a non-minimal coupling to the curvature tensor in the quantum effective action. This coupling is itself generated by quantum fluctuations even in a situation where it vanishes in the microscopic action. Finally, we draw some conclusions in section \ref{sec:Conclusions}.

\section{Symmetries and extended symmetries}
\label{sec:SymmetriesAndExtendedSymmetries}

Symmetries play an important role in quantum field theory, for at least three reasons. First, they constrain substantially the form of a microscopic action $S[\chi]$ and define in this sense to a large extend a microscopic theory. Together with renormalizability, symmetries provide the main guiding principle for the construction of microscopic physics theories.

The second reason is that symmetries constrain also to a large extend the quantum effective action $\Gamma[\phi]$ where $\phi = \langle \chi \rangle$. One may think that this is less important because $\Gamma[\phi]$ is anyway derived from $S[\chi]$ (we recall the construction below), but in practice the effective action $\Gamma[\phi]$ can usually not be obtained exactly, so that insights gained through symmetry considerations are particularly important and powerful.

In the absence of anomalies or explicit symmetry breaking, only terms that are allowed by the symmetries are allowed to appear in the quantum effective action. It is this aspect of symmetries (and extended symmetries) that will be discussed in the present section. The third important aspect of symmetries are the conservation laws they lead to. This will be discussed in the subsequent section.

While traditionally a symmetry is a transformation that leaves the action invariant, we will in the present paper also discuss more general transformations (or formally Lie group actions) for which this is not the case, but that are nevertheless very useful. Infinitesimal transformations that change the action by a term that is linear in the fields fall into this class, and they have been called extended symmetries \cite{Canet:2014cta, Tarpin:2017uzn, Tarpin:2018yvs}. We will extend this idea to transformations where the change in the action is linear in composite operators to which an external source has been coupled so that the corresponding expectation values are available on the level of the quantum effective action. However, before discussing these new developments in further detail let us recall some standard constructions, such as the quantum effective action.

\subsection{Functional integral, partition function and effective action}
We consider a theory for quantum fields $\chi(x)$ which we do not specify in further detail here. In practise $\chi(x)$ can stand for a collection of different fields and may encompass components transforming as scalars, vectors, tensors, or spinors. The theory is described by a microscopic action $S[\chi]$. The latter enters the {\it partition function},
\begin{equation}
Z[J] = \int D\chi \; e^{iS[\chi]-i \int_x \{ J(x) \chi(x)\} }.
\label{eq:SchwingerFunctional}
\end{equation}
In eq.\ \eqref{eq:SchwingerFunctional} we have introduced sources $J(x)$ for the fundamental fields $\chi(x)$. More generally one may also introduce sources for composite operators as will be discussed in more detail below. For example, the metric $g_{\mu\nu}(x)$ acts like an external source for the energy-momentum tensor $T^{\mu\nu}(x)$. We are using in \eqref{eq:SchwingerFunctional} for relativistic theories the abbreviation
\begin{equation}
\int_x = \int d^d x \sqrt{g} = \int d^d x \sqrt{-\det(g_{\mu\nu}(x))}.
\end{equation}
The space-time metric $g_{\mu\nu}(x)$ has signature $(-,+,+,+)$. 

The functional integral $\int D\chi$ is defined as usual. Even though we do not write this explicitly, a microscopic theory $S[\chi]$ is first defined in the presence of ultraviolet (and possibly infrared) regularization and for renormalizable theories there is a renormalization procedure that allows to make the corresponding ultraviolett scale arbitrarily large, in particular much larger than any other relevant mass scale. It is through this procedure that a more rigorous definition of the partition function is achieved.

It is worth to note here that the partition function in \eqref{eq:SchwingerFunctional} depends also on the quantum state or density matrix $\rho$ through the boundary conditions in the temporal domain. This is in particular important in non-trivial situations such as at finite temperature, density, or out-of-equilibrium. For initial value problems one should work with the Schwinger-Keldysh double time path formalism. For the present paper we keep the boundary condition, and therefore the state dependence, implicit. This could be easily changed, however.

From the partition function one defines the {\it Schwinger functional} $W[J] = i \ln Z[J]$ and from there the {\it quantum effective action} or {\it one-particle irreducible effective action} $\Gamma[\phi]$ as a Legendre transform,
\begin{equation}
\Gamma[\phi] =\sup_J \left( \int_x J(x) \phi(x) - W[J] \right).
\label{eq:Gammaphidef}
\end{equation}
The effective action $\Gamma[\phi]$ depends on $\phi$, which is the {\it expectation value} of the field $\chi$. To see this one evaluates the supremum by varying the source field $J(x)$ leading to
\begin{equation}
\begin{split}
\phi(x) & = \frac{1}{\sqrt{g(x)}}\frac{\delta}{\delta J(x)} W[J] = \frac{1}{\sqrt{g(x)}}\frac{i}{Z[J]} \frac{\delta}{\delta J(x)} Z[J] \\ & = \frac{\int D\chi \; \chi(x) \; e^{i S[\chi]- i \int J \chi}}{\int D\chi \; e^{iS[\chi]-i \int J \chi}}.
\end{split}
\end{equation}
One also writes this as
\begin{equation}
\phi(x) = \langle \chi(x) \rangle,
\end{equation}
with the obvious definition of the expectation value $\langle \cdot \rangle$ in the presence of sources $J$. In addition, $W[J]$ and $\Gamma[\phi]$ depend on additional sources that have beed introduced for composite operators, such as the metric $g_{\mu\nu}(x)$. 

An interesting property of $\Gamma[\phi]$ is its equation of motion. It follows from the variation of \eqref{eq:Gammaphidef} as
\begin{equation}
\frac{\delta}{\delta \phi(x)} \Gamma[\phi] = \sqrt{g(x)} J(x).
\label{eq:GammaEOM}
\end{equation}
In particular, for vanishing source $J=0$, one obtains an equation that resembles very much the classical equation of motion, $\delta S / \delta \chi =0$. However, in contrast to the latter, \eqref{eq:GammaEOM} contains all corrections from quantum fluctuations! Another interesting property is that tree-level Feynman diagrams become formally exact when propagators and vertices are taken from the effective action $\Gamma[\phi]$ instead of the microscopic action $S[\chi]$.

\subsection{Symmetry transformations of microscopic action}

Both the microscopic action $S[\chi]$ and the quantum effective action $\Gamma[\phi]$ are functionals of fields. When one speaks of a symmetry transformation of the action one means in practise a symmetry transformation of the fields on which the action depends. A symmetry of the microscopic action means an identity of the form
\begin{equation}
S[ \mathsf{g} \chi] = S[\chi].
\end{equation}
Here the group element $\mathsf{g} \in G$ is acting on the fields (not necessarily linearly) and the symmetry implies that the action is unmodified by this transformation. (More generally, the right hand side could differ from the left hand side by a constant or boundary term that does not affect the equations of motion.) So far, $G$ could be either a finite group, an infinite discrete group or a Lie group. In the latter case one can compose finite group transformations out of infinitesimal transformations. One can write for the latter
\begin{equation}
\chi(x) \to \mathsf{g} \chi(x) = \chi(x) + d \chi(x) = \chi(x) + i d\xi^j (T_j \chi) (x),
\label{eq:InfinitesimalChangeOfFields01}
\end{equation}
where $T_j$ is an appropriate representation of the Lie algebra acting on the fields $\chi$ (at this point not necessarily linearly). The microscopic action transforms as
\begin{equation}
\begin{split}
S[\mathsf{g} \chi] & = S\left[\left(\mathbbm{1} + i d\xi^j T_j \right) \chi \right] \\ & = S[\chi] +  \int d^dx \left\{ \frac{\delta S[\chi]}{\delta \chi(x)} \, i d\xi^j (T_j \chi)(x) \right\}.
\label{eq:infTransfS}
\end{split}
\end{equation}
One also abbreviate the right hand side of \eqref{eq:infTransfS} as $S[\chi] + dS[\chi]$ and a continuous symmetry corresponds then to the statement $dS[\chi]=0$.

\subsection{Symmetry transformation of integral measure}
For the transformation of the integral measure we write for an infinitesimal continuous transformation
\begin{equation}
\begin{split}
D \left(\mathsf{g} \chi \right) & = D\left(\left(\mathbbm{1} + i d\xi^j T_j \right) \chi \right)  \\ & = D \chi  \exp\left[ i \int d^dx \sqrt{g} \; i d\xi^j \mathscr{A}_j(x) \right],
\end{split}
\label{eq:anomaly}
\end{equation}
where $\mathscr{A}_j(x)$ denotes the quantum anomaly. In particular, in the absence of an anomaly one has $D \chi = D (\mathsf{g} \chi)$. (More generally one may also allow a field independent constant which can be dropped for most purposes.) 

In the following we will use a somewhat generalized notion of anomaly which also applies to transformations that are not necessarily symmetries of the microscopic action but may correspond to extended symmetries (see below). It is {\it a priori} not clear how the functional integral measure behaves with respect to these transformations or, in other words, what the relevant anomaly $\mathscr{A}_j(x)$ is. This will be an important question to be addressed in the future.

\subsection{Continuous symmetries of effective actions}

We now specialize to continuous transformations which we can study in the infinitesimal form \eqref{eq:InfinitesimalChangeOfFields01}. After a change of integration variable $\chi \to \mathsf{g} \chi$ we write the Schwinger functional \eqref{eq:SchwingerFunctional} as
\begin{equation}
\begin{split}
Z[J] = & \int D\left(\chi +i d\xi^j T_j \chi \right) \; e^{iS[\chi+i d\xi^j T_j \chi]} \\ & \times e^{- i \int_x J(x) \left(\chi(x) +i d\xi^j T_j \chi(x) \right)}.
\end{split}
\end{equation}
We use now eq.\ \eqref{eq:anomaly} and expand for small $d\xi^j$ to
\begin{equation}
\begin{split}
Z[J] = & \int D\chi \; \exp\left[iS[\chi+i d\xi^j T_j \chi] +i \int_x i d\xi^j(x) \mathscr{A}_j(x)\right] \\ & \times\exp\left[ -i \int_x J(x) \left(\chi(x) +i d\xi^j T_j \chi(x) \right)\right] \\
= & \int D\chi \; {\bigg [} 1+ 
\int_x d\xi^j {\bigg \{} {\bigg (} - \frac{1}{\sqrt{g(x)}} \frac{\delta}{\delta \chi(x)} S[\chi] \\ & + J(x) {\bigg )}  T_j \chi(x) +i  \mathscr{A}_j(x){\bigg \}} {\bigg ]} \; e^{i S[\chi] -i \int_x J(x) \chi(x)}.
\end{split}
\end{equation}
The leading term on the right hand side is just $Z[J]$ itself. Subtracting it one finds using \eqref{eq:GammaEOM} the {\it Slavnov-Taylor identity}
\begin{equation}
\begin{split}
& \left\langle dS[\chi]   + \int_x d\xi^j \mathscr{A}_j(x) \right\rangle \\ &  = \left\langle \int d^dx \left\{ \left(\frac{\delta}{\delta \chi(x)} S[\chi] \right) i d\xi^j T_j \chi(x) \right\}  + \int_x d\xi^j \mathscr{A}_j(x) \right\rangle \\
&  = \int d^dx \left\{ \left(\frac{\delta}{\delta \phi(x)} \Gamma[\phi] \right) i d\xi^j \left\langle T_j \chi(x) \right\rangle \right\}.
\end{split}
\label{eq:STIdentities}
\end{equation}
An important class of transformations is such that the Lie algebra generators $T_j$ act on the fields $\chi$ in a linear way. In that case one can write
\begin{equation}
\left\langle T_j \chi(x) \right\rangle =T_j  \left\langle \chi(x) \right\rangle = T_j \phi.
\label{eq:generatoractinglinearonfields}
\end{equation}
In that case the right hand side of \eqref{eq:STIdentities} can be written as $d\Gamma[\phi]$ and one has
\begin{equation}
\langle dS[\chi] + \int_x d\xi^j \mathscr{A}_j(x) \rangle = d \Gamma[\phi].
\label{eq:STIdentities02}
\end{equation}
In particular, the most important case is here that the microscopic action is invariant, $dS[\chi]=0$, and that the anomaly vanishes, $\mathscr{A}_j=0$, from which it follows that also the effective action is invariant, $d\Gamma[\phi]=0$, or
\begin{equation}
\Gamma[\mathsf{g}\phi] = \Gamma[\phi].
\end{equation}

In summary, in the absence of a quantum anomaly, and for a linear representation of the Lie algebra on the fields, we conclude that the effective action $\Gamma[\phi]$ shares the symmetries of the microscopic action $S[\chi]$. This is very useful in practice because it constrains very much the form the effective action can have. This is important for example for proofs of renormalizability or also for solving renormalization group equations in practice \cite{ZinnJustin:2002ru}.

\subsection{Extended symmetries} 

Interestingly, eq.\ \eqref{eq:STIdentities02} can also useful when the microscopic action $S[\chi]$ is {\it not} invariant, i.\ e.\ $dS[\chi]\neq 0$. For example, if $dS[\chi]$ is {\it linear} in the field $\chi$, one can infer that the effective action $\Gamma[\phi]$ must change under a corresponding transformation of the expectation value field $\phi$ in an analogous way such that eq.\ \eqref{eq:STIdentities02} remains fulfilled. This can also constrain the form of $\Gamma[\phi]$ substantially \cite{Canet:2014cta, Tarpin:2017uzn, Tarpin:2018yvs}. Such transformations, which are not standard symmetries because they change the action, have been called extended symmetry transformations. The fact that they are also very useful has been appreciated only rather recently.

When $dS[\chi]$ is non-linear in the fields $\chi$, eq.\ \eqref{eq:STIdentities02} is less useful to constrain the form of $\Gamma[\phi]$, because the left hand side involves then expectation values of composite operators that are not easily available on the level of the effective action $\Gamma[\phi]$. For example, the connected two-point correlation function $\langle \chi \chi \rangle - \langle \chi \rangle \langle \chi \rangle$ involves the inverse of the second functional derivative of $\Gamma[\phi]$. In such a case one may however use eq.\ \eqref{eq:STIdentities02} to calculate the expectation value on the left hand side through a simple transformation of the effective action $\Gamma[\phi]$.

Another interesting situation is when the change in the action on both sides of eq.\ \eqref{eq:STIdentities02} is linear in composite operators that are available on the level of the effective action $\Gamma[\phi]$ because external sources have been coupled to them. This is a new case, that has not been studied previously, but is is fact very interesting and useful. We will discuss several examples in section \ref{sec:Space-timeSymmetriesAndExtendedSymmetries}.

An additional example we may mention at this point is the partial conservation of the axial current in QCD, leading to the so-called PCAC relations \cite{GellMann:1960np,Nambu:1960xd,Adler:1964um,WEINBERG1979327,Gasser:1983yg,Donoghue:1992dd,Yamagishi:1995kr,Scherer:2002tk}.
In the presence of quark masses, the axial rotation of flavors is not a symmetry of the microscopic action. However one can couple 
the quark bilinear to an external field, and the quark current to an external gauge field, precisely as discussed above. One obtains then an identity for the quantum effective action which can be seen as a consequence of an extended symmetry.

Let us note here that our discussion of extended symmetries is not as extensive as one may wish it to be, specifically from a more mathematical point of view. We plan to revisit the topic in the future.

\section{Conserved and non-conserved Noether currents}
\label{sec:ConservedAndNon-conservedNoetherCurrents}
We now come to the second implication of symmetries besides constraints to the form of the microscopic and quantum effective action, namely conservation laws. We discuss two methods that can be used to extract conserved currents from the quantum effective action and will argue that the second method is superior to the first. For the first method one makes the symmetry transformations space-time dependent, while for the second method one also introduces an appropriate external gauge field. Before we go into this, let us first discuss why it is useful to obtain a Noether current from the effective action instead of the microscopic action.

\subsection{Noether current from microscopic action versus Noether current from quantum effective action}
On the classical level of a field theory one can obtain the Noether currents from the microscopic action $S[\chi]$ through the standard textbook procedure. Sometimes such an expression is useful also for a quantum description as a starting point to calculate its expectation value when quantum and statistical fluctuations are taken into account. Alternatively one can obtain the expectation value of a Noether current directly from the quantum effective action as will be discussed below.

However, one should be aware of the fact that the difference between the classical or microscopic action and the quantum effective action is a rather non-trivial result of quantum and possibly statistical fluctuations. In particular, quantum fluctuations are present on all scales and need proper regularization and renormalization procedures. It is well known that quantum fluctuations can modify a theory substantially, for example the propagating degrees of freedom can change from fundamental fields to composite fields for bound states \cite{Gies:2001nw, Pawlowski:2005xe, Floerchinger:2009uf}. In this sense it is possible that Noether currents differ substantially in form when derived from microscopic actions or quantum effective actions, respectively. Moreover, there can be contributions from anomalies to the currents that are present in the quantum effective action, but not in the microscopic or classical action. 

The above arguments show that it can be rather advantageous to calculate Noether currents directly from the quantum effective action, instead of aiming at an expression in terms of the microscopic action and expectation values in terms of complicated composite operators. 

One should note here, however, that Noether currents that follow from the quantum effective action are usually {\it not} monomials or polynomials in the field expectation values and their derivatives. There can be additional terms that do not vanish, even when the field expectation values do. For example, at finite temperature, the energy-momentum tensor is non-zero even when the field expectation values vanish. It has contributions from quantum fluctuations or quasi-particle excitations. Also, one should keep in mind that the effective action is state-dependent and accordingly the Noether currents derived from it also are. Again, the finite temperature state may serve as an example.

\subsection{Noether current from local transformations}
Consider the transformation of the fields
\begin{equation}
\phi(x) \to \phi(x) + i d\xi^j(x) \, T_j \phi(x),
\label{eq:defSymmetryTransform}
\end{equation}
where $d\xi^j(x)$ are space-time position-dependent, infinitesimal parameters. The generators $T_j$ act linearly on the field expectation value fields $\phi(x)$ such that eq.\ \eqref{eq:generatoractinglinearonfields} holds. The transformation law \eqref{eq:defSymmetryTransform} is inherited from the corresponding transformation of the microscopic fields in eq.\ \eqref{eq:InfinitesimalChangeOfFields01}. 

 The quantum effective action $\Gamma[\phi]$ changes to linear order in $d\xi^j$ like \cite{Floerchinger:2008jf}
\begin{equation}
\begin{split}
& \Gamma[\phi +i d\xi^j T_j \phi] = \Gamma[\phi] \\ &+ \int d^dx \sqrt{g} {\Big \{} \mathcal{I}_j(x) \, d\xi^j(x) + \mathcal{J^\mu}_j(x) \nabla_\mu \, d\xi^j(x) \\ & + \frac{1}{2}\mathcal{K}_j^{\mu\nu}(x) \nabla_\mu \nabla_\nu \, d\xi^j(x) + \ldots {\Big \}}.
\end{split}
\label{eq:changeInEffectiveAction}
\end{equation}
If the transformation \eqref{eq:defSymmetryTransform} is a global symmetry of the effective action this implies that $\mathcal{I}_j(x)=0$, so that the expansion on the right hand side of \eqref{eq:changeInEffectiveAction} starts with the second term, which has one derivative. However, one can also consider transformations that are not global symmetries such that $\mathcal{I}_j(x)$ is non-vanishing. One can now write eq.\ \eqref{eq:changeInEffectiveAction} after partial integration as
\begin{equation}
\begin{split}
& \int d^d x \sqrt{g} \, d\xi^j(x) {\bigg \{} \frac{1}{\sqrt{g}} \frac{\delta \Gamma}{\delta \phi(x)} i T_j \phi(x) - \mathcal{I}_j(x) \\ & + \nabla_\mu \mathcal{J}_j^\mu(x) - \frac{1}{2} \nabla_\mu \nabla_\nu \mathcal{K}_j^{\mu\nu}(x) + \ldots {\bigg \}} = 0.
\end{split}
\end{equation}
Surface terms have been dropped here. Because $d\xi^j(x)$ is arbitrary, this implies
\begin{equation}
\begin{split}
& \frac{1}{\sqrt{g}}\frac{\delta \Gamma}{\delta \phi(x)} iT_j \phi(x) - \mathcal{I}_j(x) + \nabla_\mu \mathcal{J}_j^\mu(x) \\ & - \frac{1}{2} \nabla_\mu \nabla_\nu \mathcal{K}_j^{\mu\nu}(x) + \ldots =0.
\end{split}
\end{equation}
Using the field equation \eqref{eq:GammaEOM} one obtains a set of conservation-type relations
\begin{equation}
\nabla_\mu \left( - \mathcal{J}_j^\mu(x) + \frac{1}{2} \nabla_\nu \mathcal{K}_j^{\mu\nu}(x) - \ldots \right) 
=  i J(x) T_j \phi(x) - \mathcal{I}_j(x).
\label{eq:conservationLawDerExpansion}
\end{equation}
For a transformation which defines a global symmetry such that $\mathcal{I}_j(x)=0$, and for vanishing source, $J=0$, the right hand side vanishes and the expression in brackets on the left hand side of \eqref{eq:conservationLawDerExpansion} defines then a set of conserved Noether currents. 

Let us note here that a relation of the type \eqref{eq:conservationLawDerExpansion} as derived from a quantum effective action may also be useful when $\mathcal{I}_j(x)$ is non-vanishing, as long as this field is known, for example because an external source is coupled to it. In that case one may call the expression in brackets on the left hand side of \eqref{eq:conservationLawDerExpansion} a set of {\it non-conserved Noether currents} corresponding to an {\it extended symmetry}. 

A problem with the above derivation is that the expansion on the left hand side of \eqref{eq:conservationLawDerExpansion} does in general not terminate. This makes the entire construction somewhat implicit. A notable exception is when $\Gamma[\phi] = S[\phi]$ is the microscopic or classical action which contains at most second derivatives of the fields such that the equations of motion are partial differential equations of at most second order. In this case the above construction leads to the standard Noether currents of the classical theory (the method is called Noether method in ref.\ \cite{Ortin:2015hya}). In contrast, when the effective action $\Gamma[\phi]$ differs from $S[\phi]$ by the effect of quantum fluctuations, one can not assume that only low orders of derivatives of the fields are present. In fact, the quantum effective action contains in general all orders of a derivative expansion, as well as non-perturbative terms.

These remarks show that an alternative approach is needed to obtain Noether currents from the quantum effective action when the latter cannot be assumed to have a derivative expansion terminating at finite order. To such a construction we turn next.

\subsection{Noether currents from external gauge fields}
\label{sec:NoetherCurrentsFromExternalGaugeFields}

Let us now introduce an external gauge field for the local transformation \eqref{eq:defSymmetryTransform}. All derivatives of the field $\chi(x)$ in the microscopic action and of the expectation value field $\phi(x)$ in the quantum effective action $\Gamma[\phi]$ are now replaced by covariant derivatives,
\begin{equation}
D_\mu \phi(x) = \left(\nabla_\mu - i A_\mu^j(x) T_j \right) \phi(x).
\label{eq:defCovariantDerivative}
\end{equation}
When the generators $T_j$ are not commuting, the external gauge field is non-abelian. In that case we may introduce structure constants through the relation
\begin{equation}
[T_k, T_l] = i f_{kl}^{\phantom{kl}j} T_j.
\end{equation}

The transformations of the expectation value fields $\phi(x)$ continue to be of the form \eqref{eq:defSymmetryTransform}, while the (non-abelian) gauge fields transforms as usual according to
\begin{equation}
\begin{split}
& A^j_\mu(x) \to A^j_\mu(x) + f_{kl}^{\phantom{kl}j} A_\mu^k(x) d\xi^l(x) + \nabla_\mu d\xi^j(x) \\ & = A^j_\mu(x) + (D_\mu d\xi)^j.
\end{split}
\label{eq:defSymmetryTransformGauged}
\end{equation}
In the last equation we defined the covariant derivative of $d\xi^j$ in a variant of the adjoint representation. We will call an infinitesimal transformation $d\xi^j(x)$ ``global'' when $(D_\mu d\xi)^j(x)=0$.

When the microscopic action $S[\chi, A]$ has a symmetry or extended symmetry under the transformation \eqref{eq:defSymmetryTransformGauged}, this will also be the case for the quantum effective action $\Gamma[\phi, A]$. An important consequence is that derivatives of fields $\phi$ can appear in $\Gamma[\phi, A]$ only as {\it covariant derivatives} of the form \eqref{eq:defCovariantDerivative}. One should note here, however, that $\Gamma[\phi, A]$ can contain covariant derivatives of arbitrary order, and also non-local gauge invariant terms.

The gauge field has been introduced in such a way that a local transformation (with space-time-dependent $d\xi^j(x)$) is  equivalent to a global transformation when $A^j_\mu(x)$ is transformed, as well. One has now for the transformation of the effective action
\begin{equation}
\begin{split}
& \Gamma[\phi + i d\xi^j T_j \phi, A^j_\mu + f_{kl}^{\phantom{kl}j} A_\mu^k d\xi^l + \nabla_\mu d\xi^j ]\\  & = \Gamma[\phi] + \int d^dx \sqrt{g} \left\{ \mathcal{I}_j(x) \, d\xi^j(x) \right\}.
\end{split}
\label{eq:changeInEffectiveActionLocal}
\end{equation}
In contrast to \eqref{eq:changeInEffectiveAction} no higher order derivatives are present on the right hand side of \eqref{eq:changeInEffectiveActionLocal}. This is a consequence of the external gauge field, and the fact that all derivatives have been replaced by covariant derivatives \eqref{eq:defCovariantDerivative}. Note that this includes possible regulator terms that have been added to make the functional integral well defined.\footnote{Note that the second term on the right hand side of eq.\ \eqref{eq:changeInEffectiveActionLocal} has the form of a source term for the operator $\mathcal{I}_j(x)$. Accordingly, if such a source term is already present in the setup one can possibly combine the transformations in eqs.\ \eqref{eq:defSymmetryTransform} and \eqref{eq:defSymmetryTransformGauged} with a change in this source field to obtain a standard symmetry instead of an extended symmetry.}

We can write now
\begin{equation}
\begin{split}
& \int d^d x \sqrt{g} \, {\bigg \{} \left(\frac{1}{\sqrt{g}} \frac{\delta \Gamma}{\delta \phi(x)} i T_j \phi(x) - \mathcal{I}_j(x) \right) d\xi^j(x) \\ & + \frac{1}{\sqrt{g}} \frac{\delta \Gamma}{\delta A^j_\mu(x)} \left( f_{kl}^{\phantom{kl}j} A_\mu^k(x) d\xi^l(x)  + \nabla_\mu d\xi^j(x) \right) {\bigg \}} = 0.
\end{split}
\end{equation}
Using partial integration, the field equation \eqref{eq:GammaEOM}, and the fact that $d\xi^j(x)$ is arbitrary, this implies now with the definition
\begin{equation}
\mathscr{J}_j^\mu(x) = \frac{1}{\sqrt{g}} \frac{\delta \Gamma}{\delta A^j_\mu(x)},
\end{equation}
the covariant conservation-type relation
\begin{equation}
\begin{split}
D_\mu \mathscr{J}_j^\mu(x) & = \nabla_\mu \mathscr{J}_j^\mu(x) + f_{jk}^{\phantom{jk}l} A_\mu^k(x) \mathscr{J}^\mu_l(x) \\& = i J(x) T_j \phi(x) - \mathcal{I}_j(x).
\end{split}
\label{eq:conservationLawGauged}
\end{equation}
The first equation defines a covariant derivative $D_\mu$ for the current $\mathscr{J}_j^\mu(x)$ in a variant of the adjoint representation of the gauge symmetry.

If there is a global symmetry, i.\ e.\ if the effective action is invariant for $(D_\mu d\xi)^j=0$, one has $\mathcal{I}_j(x)=0$, and for vanishing external sources, $J(x)= A_\mu^j(x)=0$, this is indeed a covariant conservation law of the standard form $\nabla_\mu \mathscr{J}_j^\mu=0$. More general, the ``source term'' on the right hand side of \eqref{eq:conservationLawGauged} might be non-vanishing. The equation is then still potentially very useful, as long as the source term is known explicitly, as for an extended symmetry transformation.

In summary, the above discussion gives a recipe to derive conservation-type equations by introducing gauge fields associated with fields transformations in the partition function and taking functional derivatives with respect to it. When the transformation corresponds to a real symmetry one obtains in this way a conserved Noether current for which all quantum corrections have been taken into account. When the transformation is instead an extended symmetry one obtains a conservation-type equation with a source term on the right hand side. It should also be seen as a macroscopic equation of motion for which quantum corrections have been taken into account. 

\section{Space-time symmetries and extended symmetries}
\label{sec:Space-timeSymmetriesAndExtendedSymmetries}
In the following we discuss a number of transformations related to space-time geometry. Following the general principles introduced in section \ref{sec:NoetherCurrentsFromExternalGaugeFields}, we introduce appropriate (external) gauge fields and discuss what kind of conservation laws follow from their variation. We start our discussion by recalling general coordinate transformations in Riemannian geometry, i.\ e.\ with the Levi-Civita connection. Subsequently we generalize this setup to a more general geometry where the connection contains additional fields beyond the Levi-Civita terms, parametrized by non-metricity and torsion. Here one can discuss local changes of frame in the tangent space of the space-time manifold. This will be done first for the restricted set of orthonormal frames where the transformations can be seen as local versions of Lorentz transformations. Subsequently we turn to general linear local frame changes which include besides local Lorentz transformations also local dilatations as well as shear transformations. We argue that the latter two should be understood as extended symmetry transformations in general. In all cases we discuss what are the associated conserved and non-conserved Noether currents.

\subsection{General coordinate transformations with the Levi-Civita connection}
\label{sec:GeneralCoordinateTransformations}
We start with a one-particle irreducible or quantum effective action $\Gamma[\phi, g]$ that depends besides the field expectation values $\phi(x)$ on the space-time metric $g_{\mu\nu}$. For the present subsection we assume that $\phi(x)$ contains only fields of integer spin or, in other words, that fermionic fields have been fully integrated out from the partition function at vanishing source. This simplifies somewhat the discussion in the sense that we do not yet have to introduce a tetrad. The matter fields $\phi(x)$ could be scalars, vectors or tensors with respect to general coordinate transformations.

Under a general coordinate transformation or diffeomorphism $x^\mu \to x^{\prime \mu}(x)$, the metric transforms like 
\begin{equation}
g_{\mu\nu}(x)\to  g^{\prime}_{\mu\nu}(x^\prime)=\frac{\partial x^{\rho}}{\partial x^{\prime \mu}} \frac{\partial x^{\sigma}}{\partial x^{\prime \nu}}  g_{\rho\sigma}(x).
\end{equation}
Changing afterwards the coordinate label from $x^{\prime\mu}$ back to $x^\mu$ gives the transformation rule
\begin{equation}
g_{\mu\nu}(x)\to  g^\prime_{\mu\nu}(x)=\frac{\partial x^{\rho}}{\partial x^{\prime \mu}} \frac{\partial x^{\sigma}}{\partial x^{\prime \nu}}  g_{\rho\sigma}(x) - \left[ g^{\prime}_{\mu\nu}(x^\prime) - g^{\prime}_{\mu\nu}(x) \right].
\end{equation}
For an infinitesimal transformation, $x^{\prime\mu}=x^{\mu} - \varepsilon^\mu(x)$, this reads
\begin{equation}
\begin{split}
g_{\mu\nu}(x) \to  & g_{\mu}(x) + \varepsilon^\rho(x) \partial_\rho g_{\mu\nu}(x) \\ & + \left(\partial_\mu\varepsilon^\rho(x) \right) g_{\rho\nu}(x) + \left(\partial_\nu\varepsilon^\rho(x) \right) g_{\mu\rho}(x) \\ & = g_{\mu\nu}(x) + \mathcal{L}_\varepsilon g_{\mu\nu}(x).
\end{split}
\label{eq:diffeoOnMetric}
\end{equation}
We are using here the {\it Lie derivative} $\mathcal{L}_\varepsilon$ in the direction $\varepsilon^\mu(x)$. More general, any coordinate tensor field transforms under infinitesimal general coordinate transformations with the corresponding Lie derivative $\mathcal{L}_\varepsilon$. This fixes in particular how the components of the matter fields $\phi(x)$ transform.

In the following we will also need the covariant derivative. In the present section it is based on the Levi-Civita connection given by the Christoffel symbols of second kind,
\begin{equation}
\Gamma_{\mu\phantom{\rho}\nu}^{\phantom{\mu}\rho} = \left\{ \small\begin{array}{c} \rho \\ \mu \nu \end{array} \right\}= \frac{1}{2} g^{\rho\lambda} \left( \partial_\mu g_{\nu\lambda} + \partial_\nu g_{\mu\lambda} - \partial_\lambda g_{\mu\nu} \right).
\label{eq:LeviCivita}
\end{equation}
For future reference we note also the variation of the Levi-Civita connection, which can be written as
\begin{equation}
\delta \left\{ \small\begin{array}{c} \rho \\ \mu \nu \end{array} \right\} = \frac{1}{2} g^{\rho\lambda} \left( \nabla_\mu \delta g_{\nu\lambda} + \nabla_\nu \delta g_{\mu\lambda} - \nabla_\lambda \delta g_{\mu\nu} \right).
\label{eq:variationConnection}
\end{equation}
The covariant derivative with notation $\nabla_\mu$ is here, and in the remainder of this paper, the covariant derivative with respect to the Levi-Civita connection in eq.\ \eqref{eq:LeviCivita}. Note that in contrast to the Christoffel symbol in \eqref{eq:LeviCivita} itself, its variation in \eqref{eq:variationConnection} is actually a coordinate tensor.

The change of the metric in eq.\ \eqref{eq:diffeoOnMetric} can also be written as
\begin{equation}
g_{\mu\nu}(x) \to g_{\mu\nu}(x) + \nabla_\mu \varepsilon_\nu(x) + \nabla_\nu \varepsilon_\mu(x).
\label{eq:variationgCovariant}
\end{equation}
This illustrates that the metric can be seen here as the gauge field of general coordinate transformations. Variation of the effective action $\Gamma[\phi, g]$ with respect to the metric at stationary matter fields, $\delta\Gamma/ \delta \phi=0$, yields the energy-momentum tensor,
\begin{equation}
\delta \Gamma[\phi, g] =  \frac{1}{2} \int d^d x \sqrt{g} \; T^{\mu\nu}(x) \delta g_{\mu\nu}(x).
\label{eq:EMTensorfrommetricVariation}
\end{equation}
In fact, $T^{\mu\nu}(x)$ as defined by this expression should be seen as the {\it expectation value} of the {\it symmetric} energy-momentum tensor (see also below) in the state that defines the effective action $\Gamma[\phi, g]$. The variation includes the connection, with \eqref{eq:variationConnection} obeyed. For an extensive discussion of eq.\ \eqref{eq:EMTensorfrommetricVariation} in the context of classical field theory see ref.\ \cite{Forger:2003ut}. 

Inserting \eqref{eq:diffeoOnMetric} in \eqref{eq:EMTensorfrommetricVariation} shows that invariance under general coordinate transformations yields the covariant conservation law for the energy momentum tensor,
\begin{equation}
\nabla_\mu T^{\mu\nu}(x) = 0.
\label{eq:standarEMConservation}
\end{equation}
In this sense, one may see the covariant conservation of energy and momentum as a special case of the general principles discussed in section \ref{sec:NoetherCurrentsFromExternalGaugeFields}.

\subsection{General connection}
\label{sec:GeneralConnection}
The Levi-Civita connection is uniquely determined by being both metric compatible and torsion free. It seems that the space-time we inhabit fulfills these two conditions to an excellent approximation. Nevertheless, it is interesting to relax these constraints and to study more general connections. Usually this is done in order to understand and constrain alternative theories of gravitation in more detail. For us the purpose is different: We are interested in constraining the form of the effective action for matter fields and to derive conservation-type relations. A very interesting possibility to this end is to study the quantum field theory in a geometry characterized by a general affine connection and to take functional derivatives of the quantum effective action with respect to the connection field. For modern introductions to non-Riemannian geometry see refs.\ \cite{Ortin:2015hya, Blagojevic:2002du, Grensing:2013leq}.

\paragraph*{Parallel transport.} As a starting point for the definition of a covariant derivative one may take the notion of a parallel transport. The rule is here that a vector field $U^\mu(x)$ counts as parallelly displaced from a position $x^\mu$ to $x^\mu+dx^\mu$ when it changes by
\begin{equation}
dU^\rho(x) = - \left[ \Gamma^{\phantom{\mu}\rho}_{\mu\phantom{\rho}\sigma}(x) - \Delta_{U}  B_\mu(x) \delta^\rho_{\phantom{\rho}\sigma} \right] U^\sigma(x) dx^\mu.
\label{eq:parallelDisplacement}
\end{equation}
The square bracket on the right hand side contains two terms. The first is a geometric part proportional to the affine connection $\Gamma^{\phantom{\mu}\rho}_{\mu\phantom{\rho}\sigma}(x)$ which generalizes the Levi-Civita connection. For the second term we take the field $U^\rho$ to have the (momentum or mass) scaling dimension or conformal weight $\Delta_{U}$. The {\it Weyl gauge field} $B_\mu(x)$ performs an additional local scaling of the field $U^\rho(x)$. In contrast to the first term in \eqref{eq:parallelDisplacement}, the second term or dilatation term is also present for scalar fields $\varphi(x)$ when they have a non-vanishing scaling dimension $\Delta_\varphi$ and when the Weyl gauge field $B_\mu(x)$ is non-vanishing.

\paragraph*{Co-covariant derivative.} The so-called co-covariant derivative 
\cite{Dirac:1973gk,Poberii:1994rz}
 associated to the parallel transport \eqref{eq:parallelDisplacement} is given by
\begin{equation}
\overline{\nabla}_\mu U^\rho(x) = \partial_\mu U^\rho(x) + \left[ \Gamma^{\phantom{\mu}\rho}_{\mu\phantom{\rho}\sigma}(x) - \Delta_{U}  B_\mu(x) \delta^\rho_{\phantom{\rho}\sigma} \right] U^\sigma(x).
\label{eq:covariantDerivativeVectorField}
\end{equation}
In particular this vanishes when $U^\rho(x)$ is parallelly transported according to \eqref{eq:parallelDisplacement}. Eq.\ \eqref{eq:covariantDerivativeVectorField} is easily generalized to other tensor fields in a coordinate basis. For example, the co-covariant derivative of a tensor field $\chi^\rho_{\;\;\lambda}(x)$ with scaling dimension $\Delta_\chi$ would be
\begin{equation}
\begin{split}
\overline{\nabla}_\mu \chi^\rho_{\;\;\lambda}(x) = & \partial_\mu \chi^\rho_{\;\;\lambda}(x) +  \Gamma^{\phantom{\mu}\rho}_{\mu\phantom{\rho}\sigma}(x) \chi^\sigma_{\;\;\lambda}(x) \\ & - \Gamma^{\phantom{\mu}\tau}_{\mu\phantom{\tau}\lambda}(x) \chi^\rho_{\;\;\tau}(x) - \Delta_{\chi}  B_\mu(x) \; \chi^\rho_{\;\;\lambda}(x).
\end{split}
\label{eq:covariantDerivativeTensorField}
\end{equation}
For a scalar field $\varphi(x)$ the co-covariant derivative is given by
\begin{equation}
\overline{\nabla}_\mu \varphi(x) = \partial_\mu \varphi(x) - \Delta_\varphi B_\mu(x) \varphi(x).
\end{equation}
The co-covariant derivative has its name because it is covariant with respect to both general coordinate transformations $x\to x^\prime(x)$ and local scaling or Weyl gauge transformations,
\begin{equation}
\begin{split}
& \phi(x) \to e^{-\Delta_\phi \zeta(x)} \phi(x), \quad\quad\quad g_{\mu\nu}(x) \to e^{2\zeta(x)} g_{\mu\nu}(x),\\ & B_\mu(x) \to B_\mu(x) - \partial_\mu \zeta(x).
\end{split}
\label{eq:WeylGaugeTransformationComplete}
\end{equation}
see below. By going to non-coordinate frames we will below also introduce variants of the co-covariant derivative \eqref{eq:covariantDerivativeTensorField} that are covariant in a generalized sense.

\paragraph*{Affine connection.} Generalizing beyond the Levi-Civita connection \eqref{eq:LeviCivita}  one may write the affine connection as
\begin{equation}
\begin{split}
\Gamma_{\mu\phantom{\rho}\sigma}^{\phantom{\mu}\rho} & =  \left\{ \small\begin{array}{c} \rho \\ \mu \sigma \end{array} \right\} + N_{\mu\phantom{\rho}\sigma}^{\phantom{\mu}\rho} \\& = \frac{1}{2} g^{\rho\lambda} \left( \partial_\mu g_{\sigma\lambda} + \partial_\sigma g_{\mu\lambda} - \partial_\lambda g_{\mu\sigma} \right) + N_{\mu\phantom{\rho}\sigma}^{\phantom{\mu}\rho},
\end{split}
\label{eq:affineConnectionDecomposition}
\end{equation}
where $N_{\mu\phantom{\rho}\sigma}^{\phantom{\mu}\rho}$ is known as the deviation or distortion tensor. (It transforms indeed as a tensor under general coordinate transformations, in contrast to $\Gamma_{\mu\phantom{\rho}\sigma}^{\phantom{\mu}\rho}$.)

\paragraph*{Co-covariant and Levi-Civita covariant derivatives.} We will use in the following a notation where $\overline{\nabla}_\mu$ denotes the co-covariant derivative based on a general affine connection as introduced in \eqref{eq:covariantDerivativeVectorField}, while $\nabla_\mu$ is the ordinary covariant derivative based on the Levi-Civita connection \eqref{eq:LeviCivita}. Equation \eqref{eq:covariantDerivativeVectorField} can also be written as
\begin{equation}
\overline{\nabla}_\mu U^\rho(x) = \nabla_\mu U^\rho(x) + \left[ N^{\phantom{\mu}\rho}_{\mu\phantom{\rho}\sigma}(x) - \Delta_{U}  B_\mu(x) \delta^\rho_{\phantom{\rho}\sigma} \right] U^\sigma(x).
\end{equation}

\paragraph*{Non-metricity.} 
The co-covariant derivative of the metric itself is given by
\begin{equation}
\begin{split}
\overline{\nabla}_\mu g_{\rho\sigma}(x) = & - \left[ N_{\mu\rho\sigma}(x) + N_{\mu\sigma\rho}(x) \right] - \Delta_{g} B_\mu(x) g_{\rho\sigma}(x). \\
= & - \left[ N_{\mu\rho\sigma}(x) + N_{\mu\sigma\rho}(x) \right] + 2 B_\mu(x) g_{\rho\sigma}(x). \\
\end{split}
\label{eq:nonMetricityCoCovariantDefinition}
\end{equation}
In the second line we have used that the metric $g_{\mu\nu}(x)$ has conformal weight $\Delta_g = -2$, as follows from eq.\ \eqref{eq:WeylGaugeTransformationComplete}. The first term on the right hand side of \eqref{eq:nonMetricityCoCovariantDefinition}, namely the combination
\begin{equation}
B_{\mu\rho\sigma}(x) = \frac{1}{2} \left[ N_{\mu\rho\sigma}(x) + N_{\mu\sigma\rho}(x) \right],
\label{eq:nonMetricityDef}
\end{equation}
is known as the non-metricity tensor. It is obviously symmetric in the last two indices. (Our convention differs by the factor $1/2$ on the right hand side of \eqref{eq:nonMetricityDef} from other places in the literature.)

It will be convenient below to further split the non-metricity tensor according to
\begin{equation}
B_{\mu\phantom{\rho}\sigma}^{\phantom{\mu}\rho}(x) = \hat B_{\mu\phantom{\rho}\sigma}^{\phantom{\mu}\rho}(x) + B_{\mu}(x) \, \delta^\rho_{\;\,\sigma},
\label{eq:splittingNonMetricity}
\end{equation}
where $\hat B_{\mu\phantom{\rho}\sigma}^{\phantom{\mu}\rho}(x) $ is trace-less and sometimes called proper non-metricity tensor, 
\begin{equation}
\hat B_{\mu\phantom{\rho}\rho}^{\phantom{\mu}\rho}(x)= 0,
\end{equation}
and $B_\mu(x) = (1/d) B_{\mu\phantom{\rho}\rho}^{\phantom{\mu}\rho}(x)$,
corresponds to the trace of the non-metricity tensor and is the {\it Weyl vector} or {\it Weyl gauge} field \cite{Weyl:1919fi} introduced already in eq.\ \eqref{eq:parallelDisplacement}. 

Note that the full co-covariant derivative of the metric in \eqref{eq:nonMetricityCoCovariantDefinition} is in fact given by the proper non-metricity tensor $-2\hat B_{\mu\rho\sigma}(x)$. 

\paragraph*{Torsion.} Consider the commutator of two co-covariant derivatives acting on a scalar field $\varphi(x)$,
\begin{equation}
\begin{split}
& \overline{\nabla}_\mu \overline{\nabla}_\nu \varphi(x) - \overline{\nabla}_\nu \overline{\nabla}_\mu \varphi(x) = - T^\rho_{\phantom{\rho}\mu\nu}(x)  \overline{\nabla}_\rho \varphi(x) \\ & - \Delta_\varphi \left[ \partial_\mu B_\nu(x) - \partial_\nu B_\mu(x) \right] \varphi(x) . 
\end{split}
\label{eq:defTorsionDerivativeCommutator}
\end{equation}
This contains two kinds of field strengths. One is the {\it torsion tensor} which is formally defined through the following combination of vector fields with vanishing scaling dimension, 
\begin{equation}
T(U,V) = \overline{\nabla}_U V - \overline{\nabla}_V U - [U, V].
\end{equation}
In components it is given by the anti-symmetric part of the affine connection,
\begin{equation}
T^\rho_{\phantom{\rho}\mu\sigma}(x) = \Gamma^{\phantom{\mu}\rho}_{\mu\phantom{\rho}\sigma}(x) - \Gamma^{\phantom{\sigma}\rho}_{\sigma\phantom{\rho}\mu}(x) = N^{\phantom{\mu}\rho}_{\mu\phantom{\rho}\sigma}(x) - N^{\phantom{\sigma}\rho}_{\sigma\phantom{\rho}\mu}(x).
\label{eq:torsionCoordinateBasis}
\end{equation}

The term in the last line of \eqref{eq:defTorsionDerivativeCommutator} is the combination $B_{\mu\nu}(x) =\partial_\mu B_\nu(x) - \partial_\nu B_\mu(x)$, known as the segmental curvature  tensor (see also below).

\paragraph*{Decomposition of distortion tensor.} Using eqs.\ \eqref{eq:nonMetricityDef} and \eqref{eq:torsionCoordinateBasis} we may write the distortion tensor as
\begin{equation}
\begin{split}
N^{\phantom{\mu}\rho}_{\mu\phantom{\rho}\sigma} = & \frac{1}{2} \left[ T^{\phantom{\mu}\rho}_{\mu\phantom{\rho}\sigma}  - T_{\sigma\mu}^{\phantom{\sigma\mu}\rho} + T^\rho_{\phantom{\rho}\mu\sigma} \right] + B^{\phantom{\mu}\rho}_{\mu\phantom{\rho}\sigma} +  B_{\sigma\mu}^{\phantom{\sigma\mu}\rho} - B^\rho_{\phantom{\rho}\mu\sigma}  \\
= &  C^{\phantom{\mu}\rho}_{\mu\phantom{\rho}\sigma} + D^{\phantom{\mu}\rho}_{\mu\phantom{\rho}\sigma}.
\end{split}
\label{eq:splittingDistortion}
\end{equation}
The combination
\begin{equation}
\begin{split}
C_{\mu\phantom{\rho}\sigma}^{\phantom{\mu}\rho} = & \frac{1}{2} \left[ T^{\phantom{\mu}\rho}_{\mu\phantom{\rho}\sigma}  - T_{\sigma\mu}^{\phantom{\sigma\mu}\rho} + T^\rho_{\phantom{\rho}\mu\sigma} \right] \\
= & \frac{1}{2} \left[ N_{\mu\phantom{\rho}\sigma}^{\phantom{\mu}\rho} - N_{\mu\sigma}^{\phantom{\mu\sigma}\rho} + N^\rho_{\phantom{\rho}\mu\sigma} + N^\rho_{\phantom{\rho}\sigma\mu} - N_{\sigma\mu}^{\phantom{\sigma\mu}\rho} - N_{\sigma\phantom{\rho}\mu}^{\phantom{\sigma}\rho} \right],
\end{split}
\end{equation}
is known as the {\it contorsion tensor}. It is anti-symmetric in the last two indices, $C_{\mu\rho\sigma} = - C_{\mu\sigma\rho}$, so it does not contribute to the non-metricity in \eqref{eq:nonMetricityDef}. In contrast, the combination
\begin{equation}
\begin{split}
D^{\phantom{\mu}\rho}_{\mu\phantom{\rho}\sigma} = & B^{\phantom{\mu}\rho}_{\mu\phantom{\rho}\sigma} +  B_{\sigma\mu}^{\phantom{\sigma\mu}\rho} - B^\rho_{\phantom{\rho}\mu\sigma} \\ = & \hat B^{\phantom{\mu}\rho}_{\mu\phantom{\rho}\sigma} +  \hat B_{\sigma\mu}^{\phantom{\sigma\mu}\rho} - \hat B^\rho_{\phantom{\rho}\mu\sigma}
+ B_\mu \delta^\rho_{\;\;\sigma} +  B_{\sigma} \delta_\mu^{\;\;\rho} - B^\rho g_{\mu\sigma}\\
= & \frac{1}{2} \left[ 
    N_{\mu\phantom{\rho}\sigma}^{\phantom{\mu}\rho} 
    + N_{\mu\sigma}^{\phantom{\mu\sigma}\rho} 
    - N^\rho_{\phantom{\rho}\mu\sigma} 
    - N^\rho_{\phantom{\rho}\sigma\mu} 
    + N_{\sigma\mu}^{\phantom{\sigma\mu}\rho} 
    + N_{\sigma\phantom{\rho}\mu}^{\phantom{\sigma}\rho} \right],
\end{split}\label{eq:conmetricity}
\end{equation}
which we may call {\it con-metricity tensor}, is symmetric in $\mu$ and $\sigma$ and does not contribute to torsion in eq.\ \eqref{eq:torsionCoordinateBasis}. 

We may therefore write the torsion tensor in terms of contorsion as
\begin{equation}
T^\rho_{\phantom{\rho}\mu\sigma}(x) = C^{\phantom{\mu}\rho}_{\mu\phantom{\rho}\sigma}(x) - C^{\phantom{\sigma}\rho}_{\sigma\phantom{\rho}\mu}(x),
\label{eq:TorsionintermsofC}
\end{equation}
and the non-metricity tensor in terms of the con-metricity tensor as
\begin{equation}
B_{\mu\rho\sigma}(x) = \frac{1}{2} \left[ D_{\mu\rho\sigma}(x) + D_{\mu\sigma\rho}(x) \right].
\end{equation}
The proper non-metricity $\hat B_{\mu\rho\sigma}(x)$ corresponds to the symmetric and trace-less part of con-metricity with respect to the last two indices. However, con-metricity has also an anti-symmetric part.
The Weyl gauge field can also be obtained directly from the trace of con-metricity as 
\begin{equation}
B_\mu(x) = \frac{1}{d} D_{\mu\phantom{\rho}\rho}^{\phantom{\mu}\rho}(x).
\end{equation}
With this, the trace of the complete affine connection \eqref{eq:affineConnectionDecomposition} can be written as
\begin{equation}
\Gamma_{\mu\phantom{\rho}\rho}^{\phantom{\mu}\rho}(x) = \frac{1}{\sqrt{g(x)}} \partial_\mu \sqrt{g(x)} + d \, B_\mu(x).
\end{equation}
In this sense the Weyl gauge field is actually determined by the affine connection and the metric,
\begin{equation}
B_\mu(x) = \frac{1}{d} \left[ \Gamma_{\mu\phantom{\rho}\rho}^{\phantom{\mu}\rho}(x) - \frac{1}{\sqrt{g(x)}} \partial_\mu \sqrt{g(x)} \right].
\label{eq:WeylGaugeFieldInTermsOfConnection}
\end{equation}
This relation underlines the geometric significance of the Weyl gauge field \cite{Weyl:1919fi}. 

For our purposes it is particularly useful to work with contorsion $C^{\phantom{\mu}\rho}_{\mu\phantom{\rho}\sigma}$, the Weyl gauge field $B_\mu$ and proper non-metricity $\hat B^{\phantom{\mu}\rho}_{\mu\phantom{\rho}\sigma}$ as the fields that parametrize the distortion tensor, so that the full connection becomes
\begin{equation}
\begin{split}
& \Gamma^{\phantom{\mu}\rho}_{\mu\phantom{\rho}\sigma} = 
\frac{1}{2} g^{\rho\lambda} \left( \partial_\mu g_{\sigma\lambda} + \partial_\sigma g_{\mu\lambda} - \partial_\lambda g_{\mu\sigma} \right) + C^{\phantom{\mu}\rho}_{\mu\phantom{\rho}\sigma} \\
& + \hat B^{\phantom{\mu}\rho}_{\mu\phantom{\rho}\sigma} +  \hat B_{\sigma\phantom{\rho}\mu}^{\phantom{\sigma}\rho} - \hat B^\rho_{\phantom{\rho}\mu\sigma}
+ B_\mu \delta^\rho_{\;\;\sigma} +  B_{\sigma} \delta_\mu^{\;\;\rho} - B^\rho g_{\mu\sigma}.
\end{split}
\label{eq:generalConnection}
\end{equation}

\paragraph*{Variation of affine connection.} The full variation of the affine connection is now given by
\begin{equation}
\begin{split}
 \delta \Gamma_{\mu\phantom{\rho}\sigma}^{\phantom{\mu}\rho} = & \frac{1}{2} g^{\rho\lambda} \left( \nabla_\mu \delta g_{\sigma\lambda} + \nabla_\sigma \delta g_{\mu\lambda} - \nabla_\lambda \delta g_{\mu\sigma} \right) + \delta N_{\mu\phantom{\rho}\sigma}^{\phantom{\mu}\rho}. %+ \delta B_\mu \delta^\rho_{\;\;\sigma} 
\end{split}
\label{eq:variationConnection2}
\end{equation}
The covariant derivative on the right hand side uses the Levi-Civita connection. 
In particular it follows from \eqref{eq:variationConnection2} that all components of the connection field $\Gamma_{\mu\phantom{\rho}\sigma}^{\phantom{\mu}\rho}$ can be varied free of constraints when this variation is understood as a superposition of the variation of the Christoffel symbols due to a variation of the metric, and variations of the torsion tensor and non-metricity tensor. In some situations one may further restrict this and demand for example that the non-metricity vanishes.

\paragraph*{Curvature tensor.}
One may define the curvature tensor by the commutator of covariant derivatives of vector fields with vanishing scaling dimension, 
\begin{equation}
\overline{R}(U,V)W=\overline{\nabla}_U \overline{\nabla}_V W - \overline{\nabla}_V \overline{\nabla}_U W - \overline{\nabla}_{[U,V]} W. 
\end{equation}
In components,
\begin{equation}
\begin{split}
\overline{R}^\rho_{\phantom{\rho}\sigma\mu\nu} = & \partial_\mu \Gamma_{\nu\phantom{\rho}\sigma}^{\phantom{\nu}\rho} - \partial_\nu \Gamma_{\mu\phantom{\rho}\sigma}^{\phantom{\mu}\rho} + 
 \Gamma_{\mu\phantom{\rho}\lambda}^{\phantom{\mu}\rho}  \Gamma_{\nu\phantom{\lambda}\sigma}^{\phantom{\nu}\lambda}  - \Gamma_{\nu\phantom{\rho}\lambda}^{\phantom{\nu}\rho}  \Gamma_{\mu\phantom{\lambda}\sigma}^{\phantom{\mu}\lambda} \\
 = & R^\rho_{\phantom{\rho}\sigma\mu\nu} + \nabla_\mu N_{\nu\phantom{\rho}\sigma}^{\phantom{\nu}\rho} - \nabla_\nu N_{\mu\phantom{\rho}\sigma}^{\phantom{\mu}\rho} \\ & + 
 N_{\mu\phantom{\rho}\lambda}^{\phantom{\mu}\rho}  N_{\nu\phantom{\lambda}\sigma}^{\phantom{\nu}\lambda}  - N_{\nu\phantom{\rho}\lambda}^{\phantom{\nu}\rho}  N_{\mu\phantom{\lambda}\sigma}^{\phantom{\mu}\lambda}.
\end{split}
 \label{eq:defRiemannTensor}
\end{equation}
This is obviously anti-symmetric in the last two indices. In the second line of \eqref{eq:defRiemannTensor}, $R^\rho_{\phantom{\rho}\sigma\mu\nu}$ is the standard Riemann tensor based on the Levi-Civita connection and the covariant derivatives are also based on the Levi-Civita connection.

It is also useful to have the variation of \eqref{eq:defRiemannTensor} at hand. It can be written as
\begin{equation}
\delta \overline{R}^\rho_{\phantom{\rho}\sigma\mu\nu} = \overline{\nabla}_\mu \delta \Gamma_{\nu\phantom{\rho}\sigma}^{\phantom{\nu}\rho} - \overline{\nabla}_\nu \delta \Gamma_{\mu\phantom{\rho}\sigma}^{\phantom{\mu}\rho} + T^\lambda_{\phantom{\lambda}\mu\nu} \, \delta \Gamma_{\lambda\phantom{\rho}\sigma}^{\phantom{\nu}\rho},
\end{equation}
with torsion as in \eqref{eq:TorsionintermsofC}. We are using here the co-covariant derivative with vanishing scaling dimension for the variation of the connection $\delta \Gamma_{\nu\phantom{\rho}\sigma}^{\phantom{\nu}\rho}$. 

\paragraph*{Ricci scalar.} There is a unique complete contraction of \eqref{eq:defRiemannTensor} which forms the analog of the Ricci scalar $R=R^{\rho\sigma}_{\phantom{\rho\sigma}\rho\sigma}$,
\begin{equation}
\begin{split}
\overline{R} = \overline{R}^{\rho\sigma}_{\phantom{\rho\sigma}\rho\sigma} = &  R + \nabla_\rho N_\sigma^{\phantom{\sigma}\rho\sigma} - \nabla_\sigma N_\rho^{\phantom{\rho}\rho\sigma} \\ & + N_{\rho\phantom{\rho}\lambda}^{\phantom{\rho}\rho} N_{\sigma}^{\phantom{\sigma}\lambda\sigma} - N_{\sigma\phantom{\rho}\lambda}^{\phantom{\rho}\rho} N_{\rho}^{\phantom{\sigma}\lambda\sigma} \\
= &  R + 2 \nabla_\rho C_\sigma^{\phantom{\sigma}\rho\sigma} + 2 \nabla_\rho \hat B_\sigma^{\phantom{\sigma}\sigma\rho} - 2(d-1) \nabla_\rho B^\rho \\ & + N_{\rho\phantom{\rho}\lambda}^{\phantom{\rho}\rho} N_{\sigma}^{\phantom{\sigma}\lambda\sigma} - N_{\sigma\phantom{\rho}\lambda}^{\phantom{\rho}\rho} N_{\rho}^{\phantom{\sigma}\lambda\sigma}.
\end{split}
\label{eq:RicciScalarAffine}
\end{equation}

In contrast to Riemann geometry, $\overline{R}_{\rho\sigma\mu\nu}$ is in general not anti-symmetric in the first two indices. Based on \eqref{eq:defRiemannTensor} one may define different contractions. One is the segmental curvature tensor
\begin{equation}
B_{\mu\nu} = \frac{1}{d} \overline{R}^\rho_{\phantom{\rho}\rho\mu\nu} = \frac{1}{d} \left[ \partial_\mu \Gamma_{\nu\phantom{\rho}\rho}^{\phantom{\nu}\rho} - \partial_\nu \Gamma_{\mu\phantom{\rho}\rho}^{\phantom{\mu}\rho} \right] = \partial_\mu B_\nu - \partial_\nu B_\mu.
\end{equation}
The other two possibilities are $\overline{R}_{\mu\nu} = \overline R^\rho_{\phantom{\rho}\mu\rho\nu}$ and $\overline R^{\phantom{\mu}\rho}_{\mu\phantom{\rho}\nu\rho}$ which both equal the standard Ricci tensor in the absence of non-metricity and torsion. 

The variation of the Ricci scalar is given by
\begin{equation}
\begin{split}
\delta \overline{R} = & - \overline{R}^{\mu\nu} \delta g_{\mu\nu} + g^{\nu\sigma} \overline{\nabla}_\rho \delta\Gamma_{\nu\phantom{\rho}\sigma}^{\phantom{\nu}\rho} - g^{\nu\sigma} \overline{\nabla}_\nu \delta\Gamma_{\rho\phantom{\rho}\sigma}^{\phantom{\nu}\rho} \\ & + T^{\lambda\phantom{\rho}\sigma}_{\phantom{\lambda}\rho}  \, \delta \Gamma_{\lambda\phantom{\rho}\sigma}^{\phantom{\lambda}\rho}.
\end{split}
\label{eq:variationRicciScalarAffine}
\end{equation}
On the right hand side one may use \eqref{eq:variationConnection2} for further simplifcation. In the absence of contorsion and non-metricity this reduces to the standard identity
\begin{equation}
\delta R = - R^{\mu\nu} \delta g_{\mu\nu} + \left[\nabla^\mu \nabla^\nu \delta g_{\mu\nu} - g^{\mu\nu} \nabla^\rho \nabla_\rho \delta g_{\mu\nu} \right].
\end{equation}

\subsection{Variation of the quantum effective action}

In the following we investigate how a quantum effective action for matter fields reacts to the contorsion and non-metrcity as external sources, and specifically what kind of equations can be derived from transformations for which the connection acts as a gauge field.

Let us write the variation of the action with respect to the metric $g_{\mu\nu}$ and the connection $\Gamma_{\mu\phantom{\rho}\sigma}^{\phantom{\mu}\rho}$ as
\begin{equation}
\begin{split}
\delta \Gamma = & \int d^d x \sqrt{g} {\bigg \{} \frac{1}{2} \mathscr{U}^{\mu\nu}(x) \delta g_{\mu\nu}(x) \\ & - \frac{1}{2} \mathscr{S}^{\mu\phantom{\rho}\sigma}_{\phantom{\mu}\rho}(x) \delta \Gamma_{\mu\phantom{\rho}\sigma}^{\phantom{\mu}\rho}(x) {\bigg \}}.
\end{split}
\label{eq:variationActionMetricConnection}
\end{equation}
The variation with respect to $g_{\mu\nu}(x)$ at fixed connection defines a symmetric tensor $\mathscr{U}^{\mu\nu}(x)$, while the variation with respect to the connection at fixed metric defines a tensor field $\mathscr{S}^{\mu\phantom{\rho}\sigma}_{\phantom{\mu}\rho}(x)$. The latter is known as {\it hypermomentum current} \cite{1976ZNatA..31..111H,1976ZNatA..31..524H,PhysRevD.17.428,Hehl:1994ue}. We have assumed in \eqref{eq:variationActionMetricConnection} that the Weyl gauge field $B_\mu(x)$ has been expressed through eq.\ \eqref{eq:WeylGaugeFieldInTermsOfConnection} in terms of the affine connection and the metric.

It is conventional and convenient to further decompose the hypermomentum current $\mathscr{S}^{\mu\phantom{\rho}\sigma}_{\phantom{\mu}\rho}(x)$ according to 
\begin{equation}
\begin{split}
\mathscr{S}^{\mu\phantom{\rho}\sigma}_{\phantom{\mu}\rho}(x) = &  Q^{\mu\phantom{\rho}\sigma}_{\phantom{\mu}\rho}(x) + W^\mu(x) \, \delta_\rho^{\phantom{\rho}\sigma} 
\\ & + S^{\mu\phantom{\rho}\sigma}_{\phantom{\mu}\rho}(x) 
+ S^{\sigma\mu}_{\phantom{\sigma\mu}\rho}(x) 
+ S^{\phantom{\rho}\mu\sigma}_{\rho}(x).
\end{split}
\label{eq:splittingHypermomentum}
\end{equation}
Here $S^{\mu\phantom{\rho}\sigma}_{\phantom{\mu}\rho}(x)$ is anti-symmetric, $S^{\mu\rho\sigma}(x)= - S^{\mu\sigma\rho}(x)$, in the last two indices and known as the {\it spin current}. It can be written in terms of the hypermomentum as
\begin{equation}
S^{\mu\rho\sigma} = \frac{1}{2} (\mathscr{S}^{\mu\rho\sigma} - \mathscr{S}^{\mu\sigma\rho}).
\label{eq:SpinCurrent}
\end{equation}
In contrast, $Q^{\mu\phantom{\rho}\sigma}_{\phantom{\mu}\rho}(x)$ is symmetric in the last two indices, $Q^{\mu\rho\sigma}(x)=Q^{\mu\sigma\rho}(x)$, and traceless, $Q^{\mu\phantom{\rho}\rho}_{\phantom{\mu}\rho}(x)=0$, and known as the (intrinsic) {\it shear current}. Finally, $W^\mu(x)$ is the (intrinsic) {\it dilatation current} or Weyl current. With these definitions we follow refs.\ \cite{1976ZNatA..31..111H, 1976ZNatA..31..524H,Hehl:1976kv}. The combination of shear current and dilatation current can be written in terms of the hypermomentum current as
\begin{equation}
\begin{split}
& Q^{\mu\rho\sigma} + W^\mu g^{\rho\sigma} = \\ &  \frac{1}{2} (
\mathscr{S}^{\mu\rho\sigma} + \mathscr{S}^{\rho\sigma\mu} 
- \mathscr{S}^{\rho\mu\sigma} + \mathscr{S}^{\mu\sigma\rho} 
+ \mathscr{S}^{\sigma\rho\mu} - \mathscr{S}^{\sigma\mu\rho}).
\end{split}
\label{eq:QWInTermsOfHyM}
\end{equation}

Alternatively to \eqref{eq:variationActionMetricConnection} one can write using \eqref{eq:variationConnection2} the variation of the quantum effective action as
\begin{equation}
\begin{split}
\delta \Gamma =  &  \int d^d x \sqrt{g} \; {\Big \{} \frac{1}{2} {\Big [} \mathscr{U}^{\mu\nu} + \frac{1}{4}\nabla_\rho ( \mathscr{S}^{\rho\mu\nu} + \mathscr{S}^{\rho\nu\mu} \\ & + \mathscr{S}^{\mu\nu\rho}+ \mathscr{S}^{\nu\mu\rho} - \mathscr{S}^{\mu\rho\nu} - \mathscr{S}^{\nu\rho\mu} ) {\Big ]} \delta g_{\mu\nu}\\
& - \frac{1}{2} \mathscr{S}^{\mu\phantom{\rho}\sigma}_{\phantom{\mu}\rho} \; \delta N_{\mu\phantom{\rho}\sigma}^{\phantom{\mu}\rho}   {\Big \}}.
\end{split}
\label{eq:GammaVariationUS}
\end{equation}
The first two lines gives the full variation of the effective action with respect to the metric at fixed distortion tensor. Because this must equal\footnote{It is in fact not fully unique how to define the energy-momentum tensor at non-vanishing distortion tensor. We use here the prescription where the variation with respect to the metric is done at fixed distortion tensor, but one could alternatively also keep, e.\ g.\ con-torsion, proper non-metricity and the Weyl gauge field fixed, which would lead to a slightly different form. For vanishing distortion tensor the definition becomes unique again.} 
\begin{equation}
\delta \Gamma = \int d^d x \sqrt{g} \;  \left\{ \frac{1}{2}T^{\mu\nu} \delta g_{\mu\nu} - \frac{1}{2} \mathscr{S}^{\mu\phantom{\rho}\sigma}_{\phantom{\mu}\rho} \delta N_{\mu\phantom{\rho}\sigma}^{\phantom{\mu}\rho} \right\},
\end{equation}
we find for the energy-momentum tensor the decomposition
\begin{equation}
\begin{split}
T^{\mu\nu} = & \mathscr{U}^{\mu\nu} + \frac{1}{4} \nabla_\rho ( \mathscr{S}^{\rho\mu\nu} + \mathscr{S}^{\rho\nu\mu} \\& + \mathscr{S}^{\mu\nu\rho}+ \mathscr{S}^{\nu\mu\rho} - \mathscr{S}^{\mu\rho\nu} - \mathscr{S}^{\nu\rho\mu}) \\
= & \mathscr{U}^{\mu\nu} + \frac{1}{2} \nabla_\rho \left( Q^{\rho\mu\nu} + W^\rho g^{\mu\nu}  \right).
\end{split}
\label{eq:energyMomentumHKdecomposition}
\end{equation}
Let us note here that $\mathscr{U}^{\mu\nu}(x)$ is not conserved by itself, even in the absense of non-metricity and torsion. The contributions from the shear current and dilatation current in \eqref{eq:energyMomentumHKdecomposition} are needed to obtain a conservation law. However, these two terms come unavoidably with derivatives.

Taking the trace of \eqref{eq:energyMomentumHKdecomposition} we obtain the divergence-type relation
\begin{equation}
\nabla_\rho W^\rho = \frac{2}{d} (T^\mu_{\phantom{\mu}\mu} - \mathscr{U}^\mu_{\phantom{\mu}\mu}).
\label{eq:divergenceW}
\end{equation}
Similarly, by subtracting the trace we find
\begin{equation}
\nabla_\rho Q^{\rho\mu\nu} = 2 (T^{\mu\nu}- \mathscr{U}^{\mu\nu}) - \frac{2}{d} (T^\sigma_{\phantom{\sigma}\sigma} - \mathscr{U}^\sigma_{\phantom{\sigma}\sigma}) g^{\mu\nu}.
\label{eq:divergenceQ}
\end{equation}
We will argue below that these two relations should be understood as conservation-type relations for non-conserved Noether currents associated to extended symmetries. While variants of eq.\ \eqref{eq:divergenceW} have been discussed in the context of dilatation and conformal symmetry (see also section \ref{sec:WeylGaugeTransformations} below), eq.\ \eqref{eq:divergenceQ} is new to the best or our knowledge.

The decomposition in \eqref{eq:energyMomentumHKdecomposition} is particularly interesting from the point of view of relativistic fluid dynamics and its derivation from quantum field theory. The first part contains the equilibrium part of the energy-momentum tensor, while the second term is by construction at least one order higher in derivatives and can give a non-equilibrium part of the energy-momentum tensor.

In the following we will investigate different transformations in the frame bundle for which the affine connection acts as a gauge field, in more detail. This will lead to further insights into the physics significance of the spin current, dilatation current and shear current. We start with Weyl transformations and local Lorentz transformations and turn then to shear transformations before we combine everything into general linear transformations.

\subsection{General coordinate transformations with general affine connection}
It is interesting to discuss the implications of general coordinate transformations or diffeomorphisms for a general affine connection with non-vanishing distortion tensor. The quantum effective action must still be a coordinate scalar, and therefore invariant under diffeomorphisms. However, it depends now on the distortion tensor as an external field, which affects the resulting conservation law. Starting from \eqref{eq:GammaVariationUS}, using the change in the metric \eqref{eq:variationgCovariant} and the change in the distortion tensor
\begin{equation}
\begin{split}
& \delta N_{\mu\phantom{\rho}\sigma}^{\phantom{\mu}\rho}  = \mathcal{L}_\varepsilon N_{\mu\phantom{\rho}\sigma}^{\phantom{\mu}\rho} \\ 
& = \varepsilon^\kappa \partial_\kappa N_{\mu\phantom{\rho}\sigma}^{\phantom{\mu}\rho} + \partial_\mu \varepsilon^\kappa N_{\kappa\phantom{\rho}\sigma}^{\phantom{\mu}\rho} - \partial_\kappa \varepsilon^\rho N_{\mu\phantom{\rho}\sigma}^{\phantom{\mu}\kappa} + \partial_\sigma \varepsilon^\kappa N_{\mu\phantom{\rho}\kappa}^{\phantom{\mu}\rho} \\
& = \varepsilon^\kappa \nabla_\kappa N_{\mu\phantom{\rho}\sigma}^{\phantom{\mu}\rho} + \nabla_\mu \varepsilon^\kappa N_{\kappa\phantom{\rho}\sigma}^{\phantom{\mu}\rho} - \nabla_\kappa \varepsilon^\rho N_{\mu\phantom{\rho}\sigma}^{\phantom{\mu}\kappa} + \nabla_\sigma \varepsilon^\kappa N_{\mu\phantom{\rho}\kappa}^{\phantom{\mu}\rho},
\end{split}
\end{equation}
one finds for the variation of the effective action
\begin{equation}
\begin{split}
\delta \Gamma = & \int d^d x \sqrt{g} \; \varepsilon^\kappa {\Big \{} - \nabla_\mu T^\mu_{\phantom{\mu}\kappa} - \frac{1}{2} \mathscr{S}^{\mu\phantom{\rho}\sigma}_{\phantom{\mu}\rho}  (\nabla_\kappa N_{\mu\phantom{\rho}\sigma}^{\phantom{\mu}\rho}) \\
& +\frac{1}{2} \nabla_\mu \left( 
\mathscr{S}^{\mu\phantom{\rho}\sigma}_{\phantom{\mu}\rho} N_{\kappa\phantom{\rho}\sigma}^{\phantom{\mu}\rho}
-\mathscr{S}^{\mu\phantom{\rho}\sigma}_{\phantom{\mu}\rho} N_{\rho\phantom{\mu}\sigma}^{\phantom{\rho}\mu}
+\mathscr{S}^{\mu\phantom{\rho}\sigma}_{\phantom{\mu}\rho} N_{\sigma\phantom{\rho}\kappa}^{\phantom{\sigma}\rho}
 \right) {\Big \}}.  
\end{split}
\end{equation}
One could insert here the decompositions for hypermomentum in eq.\ \eqref{eq:splittingHypermomentum} and the distortion tensor in eq.\ \eqref{eq:splittingDistortion} but that is not particularly instructive. The modified covariant conservation law for the energy-momentum tensor in the presence of non-metricity and torsion becomes
\begin{equation}
\begin{split}
    & \nabla_\mu T^\mu_{\phantom{\mu}\kappa} = 
    - \frac{1}{2} \mathscr{S}^{\mu\phantom{\rho}\sigma}_{\phantom{\mu}\rho}  (\nabla_\kappa N_{\mu\phantom{\rho}\sigma}^{\phantom{\mu}\rho}) \\
    & +\frac{1}{2} \nabla_\mu \left( 
    \mathscr{S}^{\mu\phantom{\rho}\sigma}_{\phantom{\mu}\rho} N_{\kappa\phantom{\rho}\sigma}^{\phantom{\mu}\rho}
    -\mathscr{S}^{\mu\phantom{\rho}\sigma}_{\phantom{\mu}\rho} N_{\rho\phantom{\mu}\sigma}^{\phantom{\rho}\mu}
    +\mathscr{S}^{\mu\phantom{\rho}\sigma}_{\phantom{\mu}\rho} N_{\sigma\phantom{\rho}\kappa}^{\phantom{\sigma}\rho} \right).
\end{split}
\label{eq:covariantConservationWithDistortionTensor}
\end{equation}
Of course, for vanishing distortion tensor this reduces to the standard covariant conservation law \eqref{eq:standarEMConservation}.

\subsection{Weyl gauge transformations}
\label{sec:WeylGaugeTransformations}
It is interesting at this point to discuss dilatations or Weyl gauge transformations as defined in eq.\ \eqref{eq:WeylGaugeTransformationComplete} in more detail. Interestingly, the general connection in eq.\ \eqref{eq:generalConnection} is left unchanged by this transformation because contributions from the Levi-Civita part and the Weyl non-metricity part cancel. This assumes that also the contorsion and proper non-metricity are invariant.

Let us also note here that $\sqrt{g} \to e^{d \zeta} \sqrt{g}$ and $T^{\mu\nu}\to e^{-(2+d) \zeta} T^{\mu\nu}$ so that the energy-momentum tensor with two upper indices has the scaling dimension $\Delta_{T^{\mu\nu}} = 2+d$. Similarly, the scaling dimension of the hypermomentum tensor must be $\Delta_{S^{\mu\;\sigma}_{\;\;\rho}} = d$.

For the effective action we find from \eqref{eq:variationActionMetricConnection} the following change under an infinitesimal Weyl transformation,
\begin{equation}
\begin{split}
\delta \Gamma & =  \int d^d x \sqrt{g} \, \frac{1}{2} \mathscr{U}^{\mu\nu}(x) \delta g_{\mu\nu}(x) \\
& = \int d^d x \sqrt{g} \, \mathscr{U}^{\mu}_{\;\;\mu}(x) \delta \zeta(x) .
\end{split}
\label{eq:changeactionWeylU}
\end{equation}
For a generic quantum field theory $\mathscr{U}^{\mu}_{\;\;\mu}(x)$ is non-vanishing and the effective action is not invariant under Weyl transformations. An exception is a scale-invariant theory at a renormalization group fixed point where $\mathscr{U}^{\mu}_{\;\;\mu}(x)=0$. (Even then there are corrections to the right hand side in curved space due to the conformal anomaly.)

It is interesting to note that a symmetry under dilatations implies $\mathscr{U}^{\mu}_{\;\;\mu}(x)=0$ and not directly a vanishing trace of the energy-momentum tensor $T^{\mu}_{\;\;\mu}(x)=0$. The latter condition would be implied by a symmetry of the theory under the larger group of conformal transformations (again up to anomalous corrections arising in curved space). 

Assume now that we consider a theory that is invariant under scaling transformations so that $\mathscr{U}^{\mu}_{\;\;\mu}(x)=0$. Equation \eqref{eq:divergenceW} implies then
\begin{equation}
T^\mu_{\;\;\mu} - \frac{d}{2} \nabla_\mu  W^\mu =0,
\label{eq:virialCurrentDivergence}
\end{equation}
In this context, $V^\mu = -\frac{d}{2} W^\mu$ is known as the virial current \cite{Jackiw:2011vz}. It was shown in ref.\ \cite{Callan:1970ze} that under the condition that the virial current is itself a divergence, $V^\mu = \nabla_\rho \sigma^{\rho\mu}$, one can actually define an ``improved'' energy-momentum tensor which is then trace-less. In practise this improvement can be done by changing the way the theory couples to space-time curvature, more specifically the Ricci scalar and Ricci tensor. In fact, it has been shown \cite{Iorio:1996ad} that if the theory has a conformal symmetry in flat space one can couple it the Ricci tensor in such as way that the energy-momentum tensor following through eq.\ \eqref{eq:EMTensorfrommetricVariation} is in fact the ``improved'' energy-momentum tensor. Assuming now that our theory is conformal and that this kind of improvement has been done implies that the energy-momentum tensor is trace-less (in flat space), and from \eqref{eq:virialCurrentDivergence} it follows that in this case
\begin{equation}
\nabla_\mu W^\mu =0.
\label{eq:conservationWeylCurrent}
\end{equation}

To summarize, the (intrinsic) dilatation or Weyl current $W^\mu$ is in general not conserved and fulfills the divergence-type relation \eqref{eq:divergenceW}. Because the right hand side is known (or calculable) one should understand $W^\mu$ as a non-conserved Noether current. For a scale-invariant system the divergence-type relation simplifies to \eqref{eq:virialCurrentDivergence}. Finally, for conformal systems one has an actual conservation law for the Weyl current \eqref{eq:conservationWeylCurrent}. One should mention here, however, that oftentimes for a conformal field theory the Weyl current actually simply vanishes, $W^\mu=0$. 

Finally, let us mention that the conservation law associated to full dilatation symmetry in Minkowski space (for a review see ref.\ \cite{Wetterich:2019qzx}) has an additional part due to the scaling of coordinates. It can be written as
\begin{equation}
J_D^\mu = x_\nu T^{\mu\nu} - \frac{d}{2} W^\mu,
\end{equation}
and is indeed conserved when eq.\ \eqref{eq:virialCurrentDivergence} is fulfilled.

\subsection{Local Lorentz transformations}
\label{sec:LocalLorentzTransformations}

As a next step we want to investigate local changes of frame that leave the space-time metric invariant. One can also understand them as a local version of Lorentz transformations. Mathematically, these transformations correspond to changes of basis in the frame bundle restricted to orthonormal frames.

Let us note that parts of the material in this subsection - the tetrad formalism for Riemannian geometry - is well known and can be found in textbooks \cite{Ortin:2015hya, Blagojevic:2002du, Grensing:2013leq, Weinberg:1972kfs}. We cover it here mainly to introduce our notation. The generalization to situations with non-vanishing non-metricity is less studied.

Orthonormal frames are anyway needed to describe fermionic fields because the standard version of the Clifford algebra uses them. (For an alternative approach see ref.\ \cite{Gies:2013noa} and references therein.) A choice of frame is usually parametrized in terms of the tetrad field, through a formalism we recall below. The tetrad can be defined formally as a Lorentz vector valued one-form $V_\mu^{\; A}(x) dx^{\mu}$. The latin index $A$ is here a Lorentz index (in a sense to be made more precise below), while the Greek index $\mu$ is a standard coordinate index. The tetrad parametrizes the change of basis in the frame bundle, and its associate bundle, from the holonomic or coordinate frame to an orthonormal frame. More precisely, $\theta^A(x) = V_\mu^{\; A}(x) dx^{\mu}$ could be seen as a new basis for one-forms, out of which any one-form can be composed, $\omega(x) = \omega_A(x) \theta^A(x)$. 

We also introduce the {\it inverse} tetrad $V_{\; \;A}^{\mu}(x)$ such that 
\begin{equation}
V^{\; A}_{\mu}(x) V_{\;\;A}^{\nu}(x) = \delta_\mu^{\;\;\nu}, \quad\quad\quad V_\mu^{\; A}(x) V^\mu_{\;\;B}(x) = \delta^A_{\;\;B}.
\end{equation}
The inverse tetrad can be seen as constituting a new basis for vectors, $v_A(x) = V_{\; \;A}^{\mu}(x) \partial_\mu$ such that any vector field can be written locally as $U(x) = U^A(x) v_A(x)$.\footnote{We are following here the conventions of ref.\ \cite{Weinberg:1972kfs}. Other authors refer to $V_{\; \;A}^{\mu}(x)$ as the tetrad field.} The dual basis for one-forms is precisely $\theta^A(x)$. 
With Minkowski metric $\eta_{AB}=\text{diag}(-1,+1,+1,+1)$ one can write the coordinate metric $g_{\mu\nu}(x)$ as
\begin{equation}
g_{\mu\nu}(x) = \eta_{AB} V^{\; A}_{\mu}(x) V^{\; B}_{\nu}(x) .
\label{eq:metricInTermsofTetrad}
\end{equation}

Under a coordinate transformation or diffeomorphism $x^\mu \to x^{\prime \mu}(x)$ on the coordinate side, the tetrad transforms like a one-form,
\begin{equation}
V^{\; A}_{\mu}(x)\to  V^{\prime A}_{\mu}(x^\prime)=\frac{\partial x^{\nu}}{\partial x^{\prime \mu}}  V^{\; A}_{\nu}(x).
\end{equation}
Changing afterwards the label or integration variable from $x^{\prime\mu}$ back to $x^\mu$ gives the transformation rule
\begin{equation}
V^{\; A}_{\mu}(x)\to  V^{\prime A}_{\mu}(x)=\frac{\partial x^{\nu}}{\partial x^{\prime \mu}}  V^{\; A}_{\nu}(x) - \left[ {V}^{\prime A}_{\mu}(x^\prime) - V^{\prime A}_{\mu}(x) \right].
\end{equation}
For an infinitesimal transformation $x^{\prime\mu}=x^{\prime\mu} - \varepsilon^\mu(x)$ this reads
\begin{equation}
\begin{split}
& V^{\; A}_{\mu}(x) \to  V^{\; A}_{\mu}(x) + \varepsilon^\nu(x) \partial_\nu V^{\; A}_{\mu}(x) + \left(\partial_\mu\varepsilon^\rho(x) \right) V^{\; A}_{\rho}(x) \\& = V^{\; A}_{\mu}(x) + \mathcal{L}_\varepsilon V^{\; A}_{\mu}(x).
\end{split}
\label{eq:diffeoOnTetrad}
\end{equation}
We are using here again the Lie derivative $\mathcal{L}_\varepsilon$ in the direction $\varepsilon^\mu(x)$. 

From \eqref{eq:metricInTermsofTetrad} and \eqref{eq:WeylGaugeTransformationComplete} one finds that under a Weyl transformation one has
\begin{equation}
V^{\; A}_{\mu}(x) \to e^{\zeta(x)} V^{\; A}_{\mu}(x),
\label{eq:WeylGaugeTransformationTetrad}
\end{equation}
so that the tetrad has the scaling dimension $\Delta_V=-1$. (Obviously the inverse tetrad has the opposite scaling dimension.)

In addition to coordinate and Weyl transformations one may also consider local Lorentz transformations or changes of the orthonormal frame acting on the tetrad according to
\begin{equation}
V^{\; A}_{\mu}(x) \to  V^{\prime A}_{\mu}(x)= \Lambda^{A}_{\;\;B}(x) \; V^{\; B}_{\mu}(x),
\label{eq:localLorentzTransform}
\end{equation}
where $\Lambda^{A}_{\;\;B}(x)$ is at every point $x$ a Lorentz transformation matrix such that
\begin{equation}
\Lambda^A_{\;\;B}(x) \Lambda^C_{\;\;D}(x) \eta_{AC} = \eta_{BD}.
\end{equation}
In other words, at every space-time point $x$ the matrices $\Lambda^A_{\;\;B}(x)$ are elements of the group SO$(1,d-1)$. Note that these local Lorentz transformations are intrinsic or {\it internal}, i.\ e.\ they do not act on the space-time argument $x$ of a field as a conventional Lorentz transformation would do.
In infinitesimal form, the local Lorentz transformation \eqref{eq:localLorentzTransform} reads
\begin{equation}
V^{\; A}_{\mu}(x) \to  V^{\prime A}_{\mu}(x) = V^{\;A}_{\mu}(x) +  d\omega^A_{\;\;B}(x) V^{\; B}_{\mu}(x),
\label{eq:infinitesimalLorentzTetrad}
\end{equation}
where $d\omega_{AB}(x)=-d\omega_{BA}(x)$ is anti-symmetric and infinitesimal. 

Coordinate vector and tensor fields can be transformed using the tetrad and its inverse to become scalars under general coordinate transformations, e.\ g.\ 
\begin{equation}
\begin{split}
& \varphi^{B}(x)= V^{\; B}_{\mu}(x) \varphi^{\mu}(x),\\ & \chi^{AB}(x)= V^{\; A}_{\mu}(x)  V^{\; B}_{\nu}(x) \chi^{\mu\nu}(x).
\end{split}
\end{equation}
The results are then Lorentz vectors and tensors, respectively. In other words these objects have now been fully transformed to the orthonormal frame. At this point it is worth to note that an action that is stationary with respect to coordinate tensor fields like $\chi^{\mu\nu}(x)$ is also stationary with respect to the resulting Lorentz tensor field $\chi^{AB}(x)$. 

More generally, a field $\Psi$ might transform in some representation $\mathcal{R}$ with respect to the local, internal Lorentz transformations or changes of orthonormal frame,
\begin{equation}
\Psi(x)\to\Psi^\prime(x) = L_\mathcal{R}(\Lambda(x)) \Psi(x),
\label{eq:transformationSpinorLocalLorentz}
\end{equation}
or infinitesimally, with Lie algebra generators $M_\mathcal{R}^{AB}$,
\begin{equation}
\Psi(x) \to \Psi^\prime(x) = \Psi(x) + \frac{i}{2} d\omega_{AB}(x) M_\mathcal{R}^{AB} \Psi(x). 
\end{equation}

One would also like to have a covariant derivative with respect to the local Lorentz transformations. 
This leads to the {\it spin connection}. The spin covariant derivative $\mathscr{D}_\mu$ is defined such that for the spinor field $\Psi(x)$ transforming under local Lorentz transformations according to \eqref{eq:transformationSpinorLocalLorentz} one has
\begin{equation}
V_{\;\; A}^{\mu}(x)\mathscr{D}_\mu \Psi (x) \to  \Lambda^{\phantom{A}B}_{A}(x) V_{\;\; B}^{\mu}(x) L_\mathcal{R}(\Lambda(x))  \mathscr{D}_\mu \Psi(x).
\end{equation}
In other words, the covariant derivative of some field transforms as before, with an additional transformation matrix for the new index, but without any extra non-homogeneous term. The full co-covariant derivative is now
\begin{equation}
\mathscr{D}_\mu  = \overline{\nabla}_{\mu} +\mathbf{\Omega}_{\mu}(x).
\label{eq:covariantDerivativeLocalLorentz}
\end{equation}
Here, $\overline{\nabla}_\mu$ is the co-covariant derivative as introduced in section \ref{sec:GeneralConnection} including the Weyl gauge field, the affine connection for coordinate indices and $\mathbf{\Omega}_{\mu}$ depends on the Lorentz representation of the field the derivative acts on. We also use the abbreviation $\mathscr{D}_A = V^\mu_{\;\;A}(x) \mathscr{D}_\mu$. 

The derivative $\mathscr{D}_\mu$ is now covariant with respect to general coordinate transformations (diffeomorphisms), Weyl gauge transformations, and local Lorentz transformations. To realize this, the spin connection $\mathbf{\Omega}_\mu(x)$ must transform like a non-abelian gauge field for local Lorentz transformations,
\begin{equation}
\begin{split}
& \mathbf{\Omega}_{\mu}(x) \to \mathbf{\Omega}_{\mu}^{\prime}(x)= L_\mathcal{R}(\Lambda(x)) \mathbf{\Omega}_{\mu}(x) L^{-1}_\mathcal{R}(\Lambda(x)) \\& -
\left[\partial_\mu L_\mathcal{R}(\Lambda(x)) \right] L^{-1}_\mathcal{R}(\Lambda(x)).
\end{split}
\end{equation}
We also write this for an infinitesimal Lorentz transformation $\Lambda^A_{\phantom{A}B}(x) = \delta^A_{\phantom{A}B} +  d\omega^A_{\;\;B}(x)$ as 
\begin{equation}
\begin{split}
\mathbf{\Omega}_{\mu}(x) \to \mathbf{\Omega}_{\mu}^{\prime}(x) =& \mathbf{\Omega}_{\mu}(x)+\frac{i}{2} d\omega_{AB}(x) \left[M_\mathcal{R}^{AB},\mathbf{\Omega}_{\mu}(x) \right] \\ & -\frac{i}{2} M_\mathcal{R}^{AB}\partial_\mu d\omega_{AB}(x).
\end{split}
\label{eq:infinitesimalLorentzSpinConnection}
\end{equation}
This is the transformation rule for a non-abelian gauge field associated to SO$(1,d-1)$. Quite generally, one may write the spin connection as
\begin{equation}
\mathbf{\Omega}_\mu(x) = \Omega_{\mu AB}(x) \frac{i}{2} M_\mathcal{R}^{AB} ,
\label{eq:defOmegaSpinConnection}
\end{equation}
where $\Omega_{\mu AB}(x)$ is anti-symmetric in the Lorentz indices $A$ and $B$ and  now independent of the representation $\mathcal{R}$. Sometimes it is also called spin connection. As an examples we note here the covariant derivative of a Lorentz vector with upper index and scaling dimension $\Delta_A$
\begin{equation}
\mathscr{D}_\mu A^B(x) = \partial_\mu A^B(x) + \Omega_{\mu\phantom{B}C}^{\phantom{\mu}B}(x) A^{C}(x) - \Delta_A B_\mu(x) A^B(x).
\end{equation}

At present, the spin connection $\Omega_{\mu\phantom{A}B}^{\phantom{\mu}A}(x)$ could be of quite general form, as long as it is anti-symmetric in the last two indices. However, in practise, it is most useful to define the spin connection $\Omega_{\mu\phantom{A}B}^{\phantom{\mu}A}$ such that the fully covariant derivative of the tetrad vanishes up to non-metricity terms,
\begin{equation}
\begin{split}
\mathscr{D}_\mu V_\nu^{\; A} & = \partial_\mu V_\nu^{\; A} + \Omega_{\mu\phantom{A}B}^{\phantom{\mu}A} V_\nu^{\; B} - \Gamma_{\mu\nu}^\rho V_\rho^{\; A} + B_\mu V_\nu^{\; A} \\
& = - \left( \hat B_{\mu\phantom{\rho}\nu}^{\phantom{\mu}\rho} + \hat B_{\nu\mu}^{\phantom{\nu\mu}\rho} - \hat B^\rho_{\phantom{\rho}\mu\nu} + B_\nu \delta_\mu^{\rho} - B^\rho g_{\mu\nu}\right) V_\rho^{\;A}.
\end{split}
\label{eq:covariantDerTetrad}
\end{equation}
Note that the terms involving non-metricity arrange such that only the Levi-Civita and torsion parts of $\Gamma^\rho_{\mu\nu}$ enter, while the non-metricity terms actually cancel. In the absence of non-metricity, eq.\ \eqref{eq:covariantDerTetrad} is known as the ``tetrad postulate'' and leads to a consistent formalism where derivatives of coordinate and Lorentz tensors are compatible.

Equation \eqref{eq:covariantDerTetrad} implies for example for some vector field $U^A$ with conformal weight $\Delta_{U^A} = \Delta_{U^\mu}-1$,
\begin{equation}
\begin{split}
V^\mu_{\;\;A}\mathscr{D}_\rho U^A = & \overline{\nabla}_\rho U^\mu - (D_{\rho\phantom{\mu}\sigma}^{\phantom{\rho}\mu} - B_\rho \delta^\mu_{\;\sigma}) T^\sigma \\
= & \nabla_\rho U^\mu + C_{\rho\phantom{\mu}\sigma}^{\phantom{\rho}\mu} U^\sigma -\Delta_{U^A} B_\rho U^\mu.
\end{split}
\end{equation}
This extends similarly for other Lorentz tensor fields. In this sense, the covariant derivative $\mathscr{D}_\mu$ acting on orthonormal frame tensor indices contains effectively a contorsion term and the Weyl gauge field, but not the proper non-metricity term.

One may solve eq.\ \eqref{eq:covariantDerTetrad} for the spin connection, leading to 
\begin{equation}
\begin{split}
\Omega_{\mu\phantom{A}B}^{\phantom{\mu}A} & = -\left(\partial_{\mu}V^{\; A}_\nu \right) V^{\nu}_{\;\;B} +  (\Gamma_{\mu\phantom{\rho}\nu}^{\phantom{\mu}\rho} -D_{\mu\phantom{\rho}\nu}^{\phantom{\mu}\rho} ) V^{\; A}_\rho V^{\nu}_{\;\; B} \\
& = - (\nabla_\mu V^{\; A}_\nu - C_{\mu\phantom{\rho}\nu}^{\phantom{\mu}\rho} V_\rho^{\;A}) V^\nu_{\;\;B} .
\end{split}
\label{eq:spinConnectionInTermsOfTetrad01}
\end{equation}
Again one observes that the non-metricity components of the affine connection cancel but contorsion remains. One can show that $\Omega_{\mu AB}$ defined by eq.\ \eqref{eq:spinConnectionInTermsOfTetrad01} is indeed anti-symmetric.

With this construction the present formalism allows to embed fermionic fields into space-times with non-vanishing torsion, Weyl gauge field, and proper non-metricity.

Finally we note a useful identity for the variation of the spin connection that can be easily derived from \eqref{eq:spinConnectionInTermsOfTetrad01}, 
\begin{equation}
\begin{split}
\delta \Omega_{\mu \phantom{A}B}^{\phantom{\mu}A}(x) = & - \left[ \mathscr{D}_\mu \delta V_\nu^{\;A} + (D_{\mu\phantom{\rho}\nu}^{\phantom{\mu}\rho} - B_\mu \delta^\rho_{\;\nu}) \right] V^\nu_{\;\; B} \\ & + \delta (\Gamma_{\mu\phantom{\rho}\nu}^{\phantom{\mu}\rho} -D_{\mu\phantom{\rho}\nu}^{\phantom{\mu}\rho} ) V_\rho^{\;A} V^\nu_{\;\;B} \\
= & - \left[ \nabla_\mu \delta V_\nu^{\;A} - C_{\mu\phantom{\rho}\nu}^{\phantom{\mu}\rho} \delta V_\rho^{\; A} + \Omega_{\mu\phantom{A}C}^{\phantom{\mu}A} \delta V_\nu^{\;C}\right] V^\nu_{\;\;B} \\
& + \delta \left( \left\{ \small\begin{array}{c} \rho \\ \mu \sigma \end{array} \right\} + C_{\mu\phantom{\rho}\sigma}^{\phantom{\mu}\rho}  \right) V_\rho^{\;A} V^\nu_{\;\;B}
.
\end{split}
\label{eq:variationSpinConnectionTetrad}
\end{equation}
We use here the fully covariant derivative $\mathscr{D}_\mu$, taking into account \eqref{eq:WeylGaugeTransformationTetrad}, and the variation of the affine connection as specified in \eqref{eq:variationConnection2} (with non-metricity canceling). Note that in contrast to the spin connection itself, which is a gauge field, its variation transforms simply as a tensor with one upper and one lower index under local Lorentz transformations. Under coordinate transformations both $\Omega_{\mu \phantom{A}B}^{\phantom{\mu}A}(x)$ and $\delta\Omega_{\mu \phantom{A}B}^{\phantom{\mu}A}(x)$ transform as one-forms.

\subsection{Conservation laws in the tetrad formalism}
\label{sec:ConservationLawsTetrad}

Let us now investigate what kinds of conservation-type relations we can obtain from the effective action $\Gamma[\phi, V, \Omega, D]$. We take the latter to depend on matter fields $\phi(x)$ which can be taken to be local Lorentz vectors, tensors and spinors. In addition the action depends on the tetrad field $V_\mu^{\; A}(x)$ which also replaces the metric everywhere. All derivatives of fields are assumed to be Lorentz-covariant derivatives $\mathscr{D}_A$ which depend on the spin connection $\Omega_{\mu}^{\phantom{\mu}AB}$.

Because of relation \eqref{eq:spinConnectionInTermsOfTetrad01}, or \eqref{eq:variationSpinConnectionTetrad}, the spin connection and can be varied independent of the tetrad and the Weyl gauge field only through a variation of contorsion. Even at vanishing physical torsion and contorsion, it is useful to consider a variation with respect to it. This is similar to varying the metric as done in section \eqref{sec:GeneralCoordinateTransformations} even though the latter is subsequently fixed, for example to describe Minkowski space.

For stationary matter fields $\delta\Gamma/\delta \phi=0$, the variation of the effective action is
\begin{equation}
\begin{split}
\delta \Gamma = &  \int d^d x \sqrt{g}  {\bigg \{} \mathscr{T}_{\phantom{\mu}A}^{\mu}(x) \delta V_{\mu}^{\;A}(x) -  \frac{1}{2}S^{\mu\phantom{A}B}_{\phantom{\mu}A}(x) \delta \Omega_{\mu\phantom{A}B}^{\phantom{\mu}A}(x)  \\&  - \left[\frac{1}{2} Q^{\mu\phantom{\rho}\sigma}_{\phantom{\mu}\rho}(x) + \frac{d}{2} W^\mu(x) \delta_\rho^{\;\;\sigma} \right] \delta D_{\mu\phantom{\rho}\sigma}^{\phantom{\mu}\rho}(x) {\bigg \}}.
\end{split}
\label{eq:variationActionTetradSpinConnection}
\end{equation}
The field $\mathscr{T}_{\phantom{\mu}A}^{\mu}(x)$ is defined through a variation with respect to the tetrad at fixed spin connection and fixed $D_{\mu\phantom{\rho}\sigma}^{\phantom{\mu}\rho}$. We will argue below that it is actually the canonical energy-momentum tensor. The variation with respect to the spin connection at fixed tetrad defines the spin current $S^{\mu\phantom{A}B}_{\phantom{\mu}A}(x)$. 

Finally the variation with respect to the proper non-metricity and Weyl gauge field at fixed tetrad and spin connection leads again to the shear current and Weyl current, respectively. All four fields $\mathscr{T}_{\phantom{\mu}A}^{\mu}(x)$, $S^{\mu}_{\phantom{\mu}AB}(x)$, $Q^{\mu\phantom{\rho}\sigma}_{\phantom{\mu}\rho}(x)$ and $W^\mu(x)$ transform under coordinate transformations and local Lorentz transformations as indicated by their indices. The reason is that the variation $\delta V_{\mu}^{\;A}(x) $, $\delta \Omega_{\mu}^{\phantom{\mu}AB}(x)$ and $\delta D_{\mu\phantom{\rho}\sigma}^{\phantom{\mu}\rho}(x)$ are all transforming as tensors in this sense and the variation of the action itself must be a scalar. 

We should also state here that a full variation of the effective action with respect to the tetrad, with the spin connection taken to obey relation \eqref{eq:spinConnectionInTermsOfTetrad01}, leads to the energy momentum tensor as a mixed coordinate and Lorentz tensor,
\begin{equation}
\begin{split}
\delta \Gamma = & \int d^d x \sqrt{g}  {\bigg \{} T^{\mu}_{\;\;A}(x) \delta V_{\mu}^{\; A}(x)  - \frac{1}{2} S^{\mu\phantom{\rho}\sigma}_{\phantom{\mu}\rho}(x) \, \delta C_{\mu\phantom{\rho}\sigma}^{\phantom{\mu}\rho}(x) \\
&  - \left[\frac{1}{2} Q^{\mu\phantom{\rho}\sigma}_{\phantom{\mu}\rho}(x) + \frac{d}{2} W^\mu(x) \delta_\rho^{\;\;\sigma} \right] \delta D_{\mu\phantom{\rho}\sigma}^{\phantom{\mu}\rho}(x) {\bigg \} }.
\end{split}
\label{eq:variationActionTetrad}
\end{equation}
Using \eqref{eq:variationSpinConnectionTetrad} we can relate the quantities in \eqref{eq:variationActionTetradSpinConnection} and \eqref{eq:variationActionTetrad} and find
\begin{equation}
\begin{split}
T^{\mu\nu}(x) = & \mathscr{T}^{\mu\nu}(x) + \frac{1}{2}\nabla_\rho \left[ S^{\rho\mu\nu}(x) 
+ S^{\mu\nu\rho}(x)
+ S^{\nu\mu\rho}(x)
 \right] \\
 & - \frac{1}{2} {\big [} 
 C_{\rho\phantom{\mu}\sigma}^{\phantom{\rho}\mu}(x) S^{\rho\nu\sigma}(x) - C_{\rho\phantom{\nu}\sigma}^{\phantom{\rho}\nu}(x) S^{\rho\mu\sigma}(x) {\big ]}.
\end{split}
\label{eq:EnergyMomentumBelinfanteRosenfeld}
\end{equation}
One can recognize the first line as the Belinfante-Rosenfeld form of the energy-momentum tensor with the first term $\mathscr{T}^{\mu\nu}(x)$ being the canonical energy-momentum tensor and $T^{\mu\nu}(x)$ its symmetric relative. The second and third line give additional terms proportional to contorsion.

Note that the expression in square brackets in the first line of \eqref{eq:EnergyMomentumBelinfanteRosenfeld} is anti-symmetric in $\rho$ and $\mu$. This implies for vanishing contorsion, 
\begin{equation}
\begin{split}
0 = \nabla_\mu T^{\mu\nu} = & \nabla_\mu \mathscr{T}^{\mu\nu} \\ & + \frac{1}{4}(\nabla_\mu\nabla_\rho - \nabla_\rho \nabla_\mu) \left[ S^{\rho\mu\nu} 
+ S^{\mu\nu\rho}
+ S^{\nu\mu\rho}
 \right]\\
= & \nabla_\mu \mathscr{T}^{\mu\nu} + \frac{1}{4} R^\nu_{\;\alpha\mu\rho} \left[  S^{\rho\mu\alpha}
+ S^{\mu\alpha\rho}
+ S^{\alpha\mu\rho}\right].
\end{split}
\label{eq:covariantConservationCanonicalEM}
\end{equation}
We have replaced in \eqref{eq:covariantConservationCanonicalEM} the commutator of covariant derivatives in terms of the standard Riemann tensor. The latter vanishes of course in flat space, so that both the symmetric and the canonical energy-momentum tensors are conserved there. However, more generally, the canonical energy-momentum tensor is conserved only up to a curvature term (and a term involving the distortion tensor if the latter is non-vanishing, see eq.\ \eqref{eq:covariantConservationWithDistortionTensor}).

In summary, the canonical energy-momentum tensor follows from a variation of the action with respect to the tetrad at fixed spin connection, while the symmetric energy-momentum tensor follows from a related variation but at contorsion kept fixed.

By construction the action is invariant under local Lorentz transformations. We consider now such a transformation in infinitesimal form. The matter fields are still assumed to be stationary, $\delta\Gamma / \delta \phi=0$, so that it suffices to consider the variations of the tetrad and spin connection,
\begin{equation}
\begin{split}
\delta \Gamma = & \int d^d x \sqrt{g} \left\{ \mathscr{T}_{\;\; A}^{\mu}(x) \delta V^{\; A}_{\mu}(x) - \frac{1}{2} S^{\mu\phantom{A}B}_{\phantom{\mu}A}(x) \delta \Omega_{\mu\phantom{A}B}^{\phantom{\mu}A}(x)\right\} \\
=  & \int d^d x \sqrt{g} {\bigg \{} \mathscr{T}_{\;\;A}^{\mu}(x) \delta\omega^A_{\;\;B}(x) V^{\; B}_{\mu}(x) \\ & - \frac{1}{2} S^{\mu\phantom{A}B}_{\phantom{\mu}A}(x) 
{\Big [} \delta\omega^A_{\;\;C}(x) \Omega_{\mu \phantom{C}B}^{\phantom{\mu}C}(x) \\ & - \Omega_{\mu \phantom{A}C}^{\phantom{\mu}A}(x) \delta\omega^C_{\;\;B}(x) - \partial_\mu \delta\omega^A_{\;\;B}(x) {\Big]}
{\bigg \}}.
\end{split}
\label{eq:variationTetratSpinConnectionIndependent}
\end{equation}
Using partial integration at vanishing non-metricity one can rewrite this as
\begin{equation}
\begin{split}
\delta \Gamma = & \int d^d x \sqrt{g} \; \delta\omega_{AB}(x) {\bigg [} \mathscr{T}^{BA}(x) - \frac{1}{2}{\Big [}\nabla_\mu S^{\mu AB}(x) \\
& +\Omega_{\mu\phantom{A}C}^{\phantom{\mu}A}(x) S^{\mu CB}(x)
+\Omega_{\mu\phantom{B}C}^{\phantom{\mu}B}(x) S^{\mu AC}(x)
{\Big ]} {\bigg ]} .
\end{split}
\end{equation}
For this to vanish for arbitrary $\delta\omega_{AB}(x)$ the expression in square brackets must be symmetric. Because $S^{\mu AB} = - S^{\mu BA}$ is anti-symmetric, we find for the divergence of the spin current
\begin{equation}
\begin{split}
& \nabla_\mu S^{\mu AB} +\Omega_{\mu\phantom{A}C}^{\phantom{\mu}A} S^{\mu CB}
+\Omega_{\mu\phantom{B}C}^{\phantom{\mu}B} S^{\mu AC} \\
& = \mathscr{T}^{BA} - \mathscr{T}^{AB}.
\end{split}
\label{eq:divergenceSpinCurrent}
\end{equation}
This is the conservation-type relation we were looking for. We argue that the spin current $S^{\mu\rho\sigma}(x)$ should be seen as a {\it non-conserved Noether current} associated to an {\it extended symmetry}. The transformation in eq.\ \eqref{eq:variationTetratSpinConnectionIndependent} is not a full symmetry in the sense of section \ref{sec:SymmetriesAndExtendedSymmetries} because a global transformation with $\mathscr{D}_\mu \delta \omega^A_{\;\;B}(x)=0$ does not make the action stationary as long as $\mathscr{T}^{AB}(x) \neq \mathscr{T}^{BA}(x)$. Nevertheless, eq.\ \eqref{eq:divergenceSpinCurrent} is still a very useful identity as long as the right hand side is known. This is indeed the case, because it follows from a variation of the quantum effective action according to eq.\ \eqref{eq:variationActionTetradSpinConnection}.

We emphasize again that the spin current is in general {\it not} conserved. What needs to be conserved as a consequence of full Lorentz symmetry in Minkowski space (also including a coordinate transformation) is the sum of spin current and orbital angular momentum current,
\begin{equation}
\mathscr{M}^{\mu AB}(x) = x^A(x) \mathscr{T}^{\mu B} (x) -  x^B(x) \mathscr{T}^{\mu A} (x) + S^{\mu AB}(x).
\end{equation}
We assume here $\mathscr{D}_\mu x^A(x) = V_\mu^{\; A}(x)$ (which essentially defines what is meant by $x^A(x)$ in non-cartesian coordinates) and one has indeed $\mathscr{D}_\mu \mathscr{M}^{\mu AB}(x)=0$ as a consequence of \eqref{eq:divergenceSpinCurrent} and the conservation law $\mathscr{D}_\mu \mathscr{T}^{\mu A} (x)=0$.

\subsection{General linear frame change transformations}
Mathematically, the frame bundle allows for change of basis transformation that are more general than the restriction to orthonormal frames we discussed above. The full group of local transformations is the general linear group GL$(d)$, which contains SO$(1,d-1)$ as a subgroup, but encompasses also dilatations and shear transformations.

We will consider the general linear group as an extension of the Lorentz group. Accordingly we introduce in addition to the generators $M^{AB}$ for infinitesimal Lorentz transformations also generators $S^{AB}$ for shear transformations and $D$ for dilatations. Note that we are using indices $A$ and $B$ as for an orthonormal frame to label the generators.

The generators for shear transformations are symmetric, $S^{AB}=S^{BA}$ and trace-less, $S^{AB} \eta_{AB}=0$. For $d$ space-time dimensions one has $d(d-1)/2$ generators $M^{AB}$, $d(d+1)/2-1$ generators $S^{AB}$ and $1$ generator $D$. Indeed, these make up the $d^2$ generators of the general linear group GL$(d)$. Without the generator for dilatations $D$, the generators $M^{AB}$ together with $S^{AB}$ generate the Lie algebra of the special linear group SL$(d)$. The fundamental represetation and the Lie brackets are recalled in appendix \ref{sec:appLieAlgebra}.

The generators of Lorentz transformation $M^{AB}$ and of dilatations $D$  each generate sub-groups, while the $S^{AB}$ alone do not. The center of the Lie algebra is generated by $D$. It will sometimes be convenient to split GL$(d)$ into to the Abelian subgroup of dilatations and the remaining group SL$(d)$.

Previously we have already discussed a group of transformations consisting of SO$(1,d-1)$ and the dilatations in terms of orthonormal frames. We will now extend first the indefinite orthogonal group SO$(1,d-1)$ to the larger group SL$(d)$ and subsequently also add the dilatation part.

Similar to the discussion of orthonormal frames in section \ref{sec:LocalLorentzTransformations} we introduce now a frame field, or soldering form, that parametrizes the change from a coordinate basis to a more general frame that we may call an {\it unimodular} frame. It can be introduced as a vector valued one form $e_\mu^{\phantom{\mu}a}(x) d x^\mu$. The smaller case latin index $a$ is now belonging to a frame that is in general neither holonomic (induced by a coordinate system), nor orthonormal. We may also introduce the inverse frame field such that
\begin{equation}
\begin{split}
e^{\; a}_{\mu}(x) e_{\;\;a}^{\nu}(x) = & \delta_\mu^{\;\;\nu},\\  e_\mu^{\; a}(x) e^\mu_{\;\;b}(x) = & \delta^a_{\;\;b}.
\end{split}
\end{equation}
The frame field $e_\mu^{\phantom{\mu}a}(x)$ behaves with respect to coordinate transformations very similar as the tetrad $V_\mu^{\phantom{\mu}A}(x)$ and we do not discuss this further.

The metric $g_{\mu\nu}(x)$ in the coordinate frame is expressed through the frame field as
\begin{equation}
\begin{split}
g_{\mu\nu}(x) = & \hat g_{ab}(x) e^{\; a}_{\mu}(x) e^{\; b}_{\nu}(x), \\ \hat g_{ab}(x) = & g_{\mu\nu}(x) e_{\; a}^{\mu}(x) e_{\; b}^{\nu}(x).
\end{split}
\label{eq:metricInTermsofFrameField}
\end{equation}
Here we introduce the metric in the {\it unimodular} frame $\hat g_{ab}(x)$. It has the property
\begin{equation}
\hat g = - \det \hat g_{ab}(x) = 1,
\end{equation}
but can otherwise be a quite general symmetric matrix. In this sense eq.\ \eqref{eq:metricInTermsofFrameField} generalizes eq.\ \eqref{eq:metricInTermsofTetrad}.

In order to discuss how general linear transformations act on the frame field, let us first exclude dilatations, which need a separate discussion because their generator is in the center of the algebra. Excluding them means here to restrict from GL$(d)$ to SL$(d)$. 
Such a special linear transformation acts on the frame field according to
\begin{equation}
e^{\; a}_{\mu}(x) \to  e^{\prime \, a}_{\mu}(x)= M^{a}_{\;\;b}(x) \; e^{\; b}_{\mu}(x),
\label{eq:localGLdTransform}
\end{equation}
where $M^{a}_{\;\;b}(x)$ is at every point $x$ a matrix with unit determinant, $M(x)\in \text{SL}(d)$. Similarly one can transform other vector fields and tensors with upper indices. Covectors and tensor fields with lower indices transform with the transpose of the inverse of $M(x)$. This makes sure that contractions of upper and lower indices can be done consistently.

Two remarks are in order 
\begin{itemize}
\item[(i)] The unimodular metric $\hat g_{ab}(x)$ and its inverse $\hat g^{ab}(x)$ are {\it not} invariant symbols with respect to SL$(d)$; the transformation law is
\begin{equation}
    \hat g_{ab}(x) \to (M^{-1})^c_{\;\;a}(x) (M^{-1})^d_{\;\;b}(x) \hat g_{cd}(x).
    \label{eq:transformationunimodularmetric}
\end{equation}

\item[(ii)] In a theory with spinor fields one would now have to work with three different frames and corresponding indices. Besides the coordinate frame and the general frame one also needs there an orthonormal frame where the Clifford algebra is rooted. An extension of spinor representations from SO$(1,d-1)$ to SL$(d)$ is not easily possible. (In principle it is possible to define the operation of general linear transformations on the Clifford algebra by employing a basis for the latter in terms of $p$-forms \cite{Floerchinger:2019oeo}, but that has substantial implications we do not discuss further here.) The transition from  the orthogonal frame to the general frame is then mediated by $e^{\;\;a}_{B}(x) = e^{\;\,a}_{\mu}(x) V^\mu_{\;\;B}(x)$. We will largely avoid this technical complication here and assume similar as in section \ref{sec:GeneralCoordinateTransformations} that all fermionic fields have been integrated out, already. We are then left with fields of integer spin that can be organized into scalar, vector and tensor representations under Lorentz transformations. These representations can be extended to general linear transformations in a rather direct way.
\end{itemize}

\paragraph*{Weyl gauge transformations of frame field.} Under a Weyl gauge transformation the frame field must transform analogously to the tetrad (see eq.\ \eqref{eq:WeylGaugeTransformationTetrad}),
\begin{equation}
e^{\; a}_{\mu}(x) \to e^{\zeta(x)} e^{\; a}_{\mu}(x).
\label{eq:WeylGaugeTransformationFrameField}
\end{equation}
Combining this with the SL$(d)$ transformation in \eqref{eq:localGLdTransform} leads to the general linear group GL$(d)$.

\paragraph*{Representations.} We consider now fields $\phi$ in some representation $\mathcal{R}$ of these generators so that an infinitesimal transformation reads
\begin{equation}
\begin{split}
\phi(x) \to \phi^\prime(x) = & \phi(x) + \frac{i}{2} d\omega_{AB}(x) M_\mathcal{R}^{AB} \phi(x) \\& + \frac{i}{2} d\zeta_{AB}(x) S_\mathcal{R}^{AB} \phi(x) + i d\zeta (x) D_\mathcal{R} \phi(x).
\end{split}
\end{equation}
As an example, a vector field $\varphi^a(x)$ is in the fundamental representation with respect to Lorentz and shear transformations and would transform for $d\zeta=0$ according to
\begin{equation}
\psi^a(x) \to \psi^{\prime a}(x) = \psi^a(x) + d\omega^a_{\phantom{a}b}(x) \psi^b(x) + d\zeta^a_{\phantom{a}b}(x) \psi^b(x).
\end{equation}
We use here $d\omega^a_{\phantom{a}b}(x) =d\omega^A_{\phantom{A}B}(x) e_A^{\;\;a}(x) e^B_{\;\;b}(x)$ etc.  Note that Lorentz boosts parametrized by $d\omega^a_{\phantom{a}b}(x)$ and shear transformations parametrized by $d\zeta^a_{\phantom{a}b}(x)$ are represented in a closely related way. 

More formally,  the fundamental representation has the generators
\begin{equation}
\begin{split}
(M_\mathcal{F}^{AB}(x))^c_{\;\;d} = & -i \left[ e^{Ac}(x) e^B_{\;\;d}(x) - e^{Bc}(x) e^A_{\;\;d}(x) \right], \\
(S_\mathcal{F}^{AB}(x))^c_{\;\;d} = & -i \left[ e^{Ac}(x) e^B_{\;\;d}(x) + e^{Bc}(x) e^A_{\;\;d}(x) \right.\\ & \left. - (2/d) \eta^{AB} \delta^c_{\;\;d} ]\right. .
\end{split}
\end{equation}
Note that the generators depend here on the space-time position $x$. In a similar way one can find other tensor representations. For example, a covector field would transform as
\begin{equation}
\chi_b(x) \to \chi^{\prime}_b(x) = \chi_b(x) - d\omega^a_{\phantom{a}b}(x) \chi_a(x) - d\zeta^a_{\phantom{a}b}(x) \chi_a(x).
\end{equation}
A general $(n,m)$-tensor representation of SL$(d)$ changes under a finite group transformation as
\begin{equation}
\begin{split}
& \phi^{a_1 \cdots a_n}_{b_1 \cdots b_m}(x) \to M^{a_1}_{\phantom{a_1}c_1}(x) \cdots M^{a_n}_{\phantom{a_1}c_n}(x)  \\& \times (M^{-1})^{d_1}_{\phantom{d_1}b_1}(x) \cdots (M^{-1})^{d_m}_{\phantom{d_m}b_m}(x) \; \phi^{c_1 \cdots c_n}_{d_1 \cdots d_m}(x).
\end{split}
\end{equation}

\paragraph*{Dilatations.} Let us now come to dilatations. Because they are in the center of the algebra one can assign in principle an arbitrary charge to some field $\varphi(x)$. Usually this is done such that a scalar field with the (momentum) scaling dimension $\Delta_\phi$ would transform as in eq.\ \eqref{eq:WeylGaugeTransformationComplete}.
In a similar way, any $(n,m)$-tensor field in an orthonormal frame would transform under dilatations according to its scaling dimension $\Delta_\phi$,
\begin{equation}
\phi^{A_1 \cdots A_n}_{B_1 \cdots B_m}(x) \to 
\exp(-\zeta(x) \Delta_\phi) \; \phi^{A_1 \cdots A_n}_{B_1 \cdots B_m}(x).
\label{eq:transformphiscaling}
\end{equation}

Dilatations in the unimodular frame are as in an orthonormal frame, because the transition matrix $e^B_{\;\;a}(x) = e^\mu_{\;\;a}(x) V_\mu^{\;\;B}(x)$ has the scaling dimension $\Delta_{e^B_{\;\;a}} = 0$. This implies in particular that the metric in the unimodular frame $\hat g_{ab}(x)$ also has vanishing scaling dimension and for a general tensor one has
\begin{equation}
\phi^{a_1 \cdots a_n}_{b_1 \cdots b_m}(x) \to 
\exp(- \zeta(x) \Delta_\phi ) \; \phi^{a_1 \cdots a_n}_{b_1 \cdots b_m}(x).
\label{eq:transformphiscaling2}
\end{equation}

\paragraph*{Covariant derivative.} In order to make derivatives transform in the appropriate representation of GL$(d)$, we need to define an appropriately generalized covariant derivative. We will write the latter as 
\begin{equation}
\mathcal{D}_\mu = \overline{\nabla}_\mu + \mathbf{\Omega}_\mu(x).
\end{equation}
Equation \eqref{eq:defOmegaSpinConnection} is now generalized to
\begin{equation}
\mathbf{\Omega}_\mu(x) = \Omega_{\mu AB}(x)\left( \frac{i}{2} M_\mathcal{R}^{AB}(x) + \frac{i}{2} S_\mathcal{R}^{AB}(x) + \frac{i}{d}  \eta^{AB} D_\mathcal{R} \right) .
\end{equation}
The general linear connection $ \Omega_{\mu AB}(x)$ is now not anti-symmetric as the spin connection in the last two indices anymore, but has also a symmetric and trace-less contribution which determines the shear transformation sector, as well as a trace which governs dilatations. The trace is directly related to the Weyl gauge field by
\begin{equation}
B_\mu(x) = \frac{1}{d} \Omega_{\mu\phantom{a}a}^{\phantom{\mu}a}(x).
\label{eq:WeylGaugeFieldThroughOmega}
\end{equation}

For tensor representations of GL$(d)$ one can directly work with
\begin{equation}
\Omega_{\mu \phantom{a}b}^{\phantom{\mu}a}(x) = \Omega_{\mu AB}(x) e^{Aa}(x) e^{B}_{\;\;b}(x),
\end{equation}
so that for example $\mathcal{D}_\mu \chi^a_{\;\;b}(x) = \partial_\mu  \chi^a_{\;\;b}(x) + \Omega_{\mu \phantom{a}c}^{\phantom{\mu}a}(x) \chi^c_{\;\;b}(x) - \Omega_{\mu \phantom{c}b}^{\phantom{\mu}c}(x) \chi^a_{\;\;c}(x) - B_\mu(x) \Delta_\chi \chi^a_{\;\;b}(x)$.

\paragraph*{Gauge transformations.} The general linear connection is now a gauge field for the group GL$(d)$. As such, it transforms as
\begin{equation}
\begin{split}
\Omega_{\mu \phantom{a}b}^{\phantom{\mu}a}(x) \to \Omega_{\mu \phantom{a}b}^{\prime\;a}(x)  = & M^a_{\;\;c}(x)   \Omega_{\mu \phantom{c}d}^{\phantom{\mu}c}(x) (M^{-1})^d_{\;\;b}(x) \\&  - \left[ \partial_\mu M^a_{\;\;c}(x) \right] (M^{-1})^c_{\;\;b}(x).
\end{split}
\label{eq:GLConnectionGaugeTransform}
\end{equation}
For an infinitesimal transformation this reads
\begin{equation}
\begin{split}
& \Omega_{\mu \phantom{a}b}^{\phantom{\mu}a}(x) \to \Omega_{\mu \phantom{a}b}^{\prime\;a}(x) \\  
= & \Omega_{\mu \phantom{a}b}^{\phantom{\mu}a}(x)  + d\omega^a_{\;\;c}(x) \Omega_{\mu \phantom{c}b}^{\phantom{\mu}c}(x) - \Omega_{\mu \phantom{a}c}^{\phantom{\mu}a}(x) d\omega^c_{\;\;b}(x) \\&  - \partial_\mu d\omega^a_{\;\;b}(x) + d\zeta^a_{\;\;c}(x) \Omega_{\mu \phantom{c}b}^{\phantom{\mu}c}(x) - \Omega_{\mu \phantom{a}c}^{\phantom{\mu}a}(x) d\zeta^c_{\;\;b}(x) \\& - \partial_\mu d\zeta^a_{\;\;b}(x) - \partial_\mu d\zeta(x) \delta^a_{\;\;b}\\
= & \Omega_{\mu \phantom{a}b}^{\phantom{\mu}a}(x)  -  \mathcal{D}_\mu (d\omega^a_{\;\;b} + d\zeta^a_{\;\;b} + d\zeta \delta^a_{\;\;b}).
\end{split}
\label{eq:transformationGeneralizedSpinConnection}
\end{equation}

Similar to the spin connection, the general linear connection $\Omega_{\mu\phantom{a}b}^{\phantom{\mu}a}(x)$ should be defined such that the fully covariant derivative of the frame field vanishes,
\begin{equation}
\mathcal{D}_\mu e_\nu^{\; a} = \partial_\mu e_\nu^{\; a} + \Omega_{\mu\phantom{a}b}^{\phantom{\mu}a} e_\nu^{\; b} - \Gamma_{\mu\nu}^\rho e_\rho^{\; a} = 0.
\label{eq:covariantDivergenceTetradGeneralized}
\end{equation}
Note that due to eq.\ \eqref{eq:WeylGaugeFieldThroughOmega} this also naturally contains the Weyl gauge field with the right prefactor.

Eq.\ \eqref{eq:transformationGeneralizedSpinConnection} has the advantage that derivatives of tensors can be consistently evaluated in terms of the coordinate or the general linear frame, but some care is needed concerning the conformal scaling weight. For example, for a vector field in a unimodular frame $U^a$ with conformal scaling weight $\Delta_{U}$
\begin{equation}
\begin{split}
e^\rho_{\;\;a} \mathcal{D}_\mu U^a = & e^\rho_{\;\;a} {\Big [} \partial_\mu U^a + \left( \Omega_{\mu\phantom{a}b}^{\phantom{\mu}a} - \Omega_{\mu\phantom{c}c}^{\phantom{\mu}c} \delta^a_{\;\;b} / d \right) U^b  \\
& - (\Delta_{U}/d) \Omega_{\mu\phantom{\mu}c}^{\phantom{\mu}c} U^a {\Big ]} \\
= & \partial_\mu U^\rho + \Gamma_{\mu\phantom{\rho}\sigma} U^\sigma - (\Delta_{U}+1) B_\mu U^\rho \\
= & \overline{\nabla}_\mu U^\rho,
\end{split}
\end{equation}
where the conformal scaling weight in the coordinate frame is $\Delta_U +1$ as expected.

One may solve eq.\ \eqref{eq:covariantDivergenceTetradGeneralized} for the general linear connection, leading to
\begin{equation}
\Omega_{\mu\phantom{a}b}^{\phantom{\mu}a} 
= -\left(\partial_{\mu}e^{\; a}_\nu \right) e^{\nu}_{\;\;b} +  \Gamma^\rho_{\mu\nu}e^{\; a}_\rho e^{\nu}_{\;\; b} = -\left(\overline{\nabla}_{\mu}e^{\; a}_\nu \right) e^{\nu}_{\;\;b}.
\label{eq:spinConnectionInTermsOfTetradGeneralized01}
\end{equation} 
One may check that \eqref{eq:spinConnectionInTermsOfTetradGeneralized01} transforms also correctly, i.\ e.\ according to \eqref{eq:transformationGeneralizedSpinConnection}, and that in contrast to the spin connection \eqref{eq:spinConnectionInTermsOfTetrad01}, $\Omega_{\mu ab}(x)$ is now {\it not} anti-symmetric in $a$ and $b$ any more. Also, with eq.\ \eqref{eq:WeylGaugeFieldInTermsOfConnection} one can see that eq.\ \eqref{eq:WeylGaugeFieldThroughOmega} in indeed fulfilled.

An equation analogous to \eqref{eq:variationSpinConnectionTetrad} also holds for the variation $\delta \Omega_{\mu\phantom{a}b}^{\phantom{\mu}a}(x)$,
\begin{equation}
\delta \Omega_{\mu \phantom{a}b}^{\phantom{\mu}a}(x) = - \left( \mathcal{D}_\mu \delta e_\nu^{\;a} \right) e^\nu_{\;\; b} + \delta \Gamma^\rho_{\mu\nu} e_\rho^{\;\,a} e^\nu_{\;\;b}.
\label{eq:variationGLConnection}
\end{equation}
One may use here \eqref{eq:variationConnection2} for the variation $\delta\Gamma^{\rho}_{\mu\nu}$. It is important to note here that the general linear connection \eqref{eq:GLConnectionGaugeTransform} is consistently defined also with non-vanishing contorsion $C_{\mu\phantom{\rho}\sigma}^{\phantom{\mu}\rho}(x)$ and non-metricity $B_{\mu\phantom{\rho}\sigma}^{\phantom{\mu}\rho}(x)$. In this sense the variation $\delta\Omega_{\mu\phantom{a}b}^{\phantom{\mu}a}(x)$ in \eqref{eq:variationGLConnection} is free of algebraic constraints, even if the frame field $e_\nu^{\;\; a}(x)$ is kept fixed.

From eqs.\ \eqref{eq:metricInTermsofFrameField}, \eqref{eq:covariantDivergenceTetradGeneralized} and \eqref{eq:nonMetricityDef} one finds
\begin{equation}
\mathcal{D}_\mu \hat g_{ab} = \partial_\mu \hat g_{ab} - \Omega_{\mu ab} - \Omega_{\mu ba} + \frac{2}{d} \Omega_{\mu\phantom{c}c}^{\phantom{\mu}c} \hat g_{ab} = -2 \hat B_{\mu ab}.
\label{eq:cocovariantDerivunimodularmetric}
\end{equation}
The unimodular metric $\hat g_{ab}(x)$ is only covariantly constant for vanishing proper non-metricity.

\paragraph*{Cartan's structure equations.} For completeness we also note Cartan's first structure equation for torsion,
\begin{equation}
T_{\phantom{a}\mu\nu}^{a} = \partial_\mu e^{\; a}_{\nu} - \partial_\nu e^{\; a}_{\mu} + \Omega_{\mu\phantom{a}b}^{\phantom{\mu}a} e^{\; b}_{\nu} - \Omega_{\nu\phantom{a}b}^{\phantom{\nu}a} e^{\;b}_{\mu}.
\label{eq:CartanFirstStructureEquation}
\end{equation}
Using eq.\ \eqref{eq:covariantDivergenceTetradGeneralized} one can see that this is indeed in agreement with eq.\ \eqref{eq:torsionCoordinateBasis}. In particular the right hand side vanishes in situations without space-time torsion. Similarly, Cartan's second structure equation yields the curvature tensor,
\begin{equation}
\overline{R}^a_{\phantom{a}b\mu\nu} = \partial_\mu \Omega_{\nu\phantom{a}b}^{\phantom{\nu}a} - \partial_\nu \Omega_{\mu\phantom{a}b}^{\phantom{\nu}a} +  \Omega_{\mu\phantom{a}c}^{\phantom{\nu}a} \Omega_{\nu\phantom{c}b}^{\phantom{\nu}c} - \Omega_{\nu\phantom{a}c}^{\phantom{\nu}a} \Omega_{\mu\phantom{c}b}^{\phantom{\nu}c}.
\label{eq:CartanRiemann}
\end{equation}
The Ricci scalar is given by $\overline{R} = g^{\sigma\nu} e^\mu_{\;\;a} e_\sigma^{\;\;b} \overline{R}^a_{\phantom{a}b\mu\nu}$.

\paragraph*{Some possible choices for the frame field.} The unimodular frame field $e_\mu^{\phantom{\mu}a}(x)$ can be left open, but it can also be fixed in different ways. Two possibilities are particularly interesting.
\begin{enumerate}
\item An orthonormal frame is a special case of an unimodular frame. This is obtained by fixing $\hat g_{ab}(x) = \eta_{ab}$ to the Minkowski metric. The frame field is then a tetrad field, with all the properties discussed in sections \ref{sec:LocalLorentzTransformations} and \ref{sec:ConservationLawsTetrad}. Note, however, that the connections in the orthogonal frame $\Omega_{\mu\phantom{A}B}^{\phantom{\mu}A}$ and in the unimodular frame $\Omega_{\mu\phantom{a}b}^{\phantom{\mu}a}$ have been defined differently, and they have different algebraic properties.
\item One can also set $e_\mu^{\phantom{\mu}a}(x) = \delta_\mu^{\phantom{\mu}a} / \chi(x) $. Here we have introduced a kind of external {\it dilaton field} $\chi(x)$ with scaling dimension $\Delta_\chi=1$ such that it transforms under Weyl transformations according to $\chi(x) \to e^{-\zeta(x)} \chi(x)$. Accordingly the frame field has the correct scaling dimension. For this choice the coordinate metric is of the form $g_{\mu\nu}(x) = \hat g_{\mu\nu}(x) / \chi(x)^2$. Because of $\hat g = - \det \hat g_{ab}(x) =1$ one has $g(x) = - \det g_{\mu\nu}(x) = \chi(x)^{-2d}$.  
\end{enumerate}

\subsection{Conservation laws in a general linear frame}
Let us now discuss the response of a quantum field theory to general linear changes of frame. We will again employ the quantum effective action which depends on matter field expectation values $\phi(x)$, the frame field $e_\mu^{\;\,a}(x)$ and the general linear connection $\Omega_{\mu\phantom{a}b}^{\phantom{\mu}a}(x)$. In addition it also depends on the metric in the unimodular frame $\hat g_{ab}(x)$. Because the metric has fixed determinant, $\hat g = - \det \hat g_{ab}(x) = 1$, its variation is trace-free, $\hat g^{ab}(x) \delta \hat g_{ab}(x)=0$.

Because of \eqref{eq:variationGLConnection} and \eqref{eq:variationConnection2} the general linear connection and the frame field $e_\mu^{\;a}(x)$ are only independent when contorsion and non-metricity are allowed to vary. We also need to carry the metric $\hat g_{ab}(x)$ because it is needed to construct the coordinate metric $g_{\mu\nu}(x)$ according to eq.\ \eqref{eq:metricInTermsofFrameField}. Also, $\hat g_{ab}$ and its inverse may of course appear in the effective action.

We write the effective action as
\begin{equation}
\Gamma[\phi, e, \Omega, \hat g],
\end{equation}
and for stationary matter fields it has the variation
\begin{equation}
\begin{split}
\delta \Gamma = \int d^d x \sqrt{g} & {\bigg \{} \mathscr{T}_{\phantom{\mu}a}^{\mu}(x) \delta e_{\mu}^{\;\,a}(x) -  \frac{1}{2}\mathscr{S}^{\mu\phantom{a}b}_{\phantom{\mu}a}(x) \delta \Omega_{\mu\phantom{a}b}^{\phantom{\mu}a}(x)  \\ & + \frac{1}{2} \hat{ \mathscr{U}}^{ab}(x) \delta \hat g_{ab}(x) {\bigg \}}.
\end{split}
\label{eq:variationActionTetradGeneralizedSpinConnection}
\end{equation}
This defines a field $\mathscr{T}_{\phantom{\mu}a}^{\mu}$ which must be the canonical energy-momentum tensor in the unimodular frame. This becomes clear when one compares with \eqref{eq:variationActionTetradSpinConnection} and realizes that $\hat g_{ab}$ could be kept fixed at the Minkowski form $\eta_{ab}$, such that the frame field $e_{\mu}^{\;\,a}$ is just the tetrad. 

Similarly, the field $\mathscr{S}^{\mu\phantom{a}b}_{\phantom{\mu}a}$ must be the the {\it hypermomentum current} \cite{1976ZNatA..31..111H,1976ZNatA..31..524H,PhysRevD.17.428,Hehl:1994ue} introduced already in eq.\ \eqref{eq:variationActionMetricConnection}, now in the unimodular frame. This is because keeping the frame field $e_{\mu}^{\;\,a}$ and $\hat g_{ab}$ fixed means to keep also the coordinate metric $g_{\mu\nu}$ fixed and according to \eqref{eq:variationGLConnection} one has then $\delta \Omega_{\mu\phantom{a}b}^{\phantom{\mu}a} = \delta \Gamma^\rho_{\mu\nu} e_\rho^{\;\,a} e^\nu_{\;\;b}$.  

Finally, $\hat{\mathscr{U}}^{ab}$ must be the trace-less part of the field $\mathscr{U}^{\mu\nu}(x)$ introduced in \eqref{eq:variationActionMetricConnection}, now in the unimodular frame,
\begin{equation}
\hat{\mathscr{U}}^{ab} = \left[\mathscr{U}^{\mu\nu} - (1/d) g^{\mu\nu} \mathscr{U}^\rho_{\phantom{\rho}\rho}\right] e_\mu^{\;\;a} e_\nu^{\;\;b}.
\end{equation}
This is because keeping the frame field $e_{\mu}^{\;\,a}$ and the connection $\Omega_{\mu\phantom{a}b}^{\phantom{\mu}a}$ fixed but varying the metric $\hat g_{ab}$ is like keeping the connection $\Gamma_{\mu\phantom{\rho}\sigma}^{\phantom{\mu}\rho}$ fixed and varying only the metric such that $g^{\mu\nu} \delta g_{\mu\nu} =0$.

If instead the variation is done such that \eqref{eq:variationGLConnection} is obeyed, we write
\begin{equation}
\begin{split}
\delta \Gamma = \int d^d x \sqrt{g}  {\Big \{}  T^{\mu}_{\;\;a}(x) \delta e_{\mu}^{\; a}(x) + \frac{1}{2} \hat T^{ab}(x) \delta \hat g_{ab}(x) \\
- \frac{1}{2} \mathscr{S}^{\mu\phantom{\rho}\sigma}_{\phantom{\mu}\rho}(x) \delta N_{\mu\phantom{\rho}\sigma}^{\phantom{\mu}\rho}(x)
{\Big \} }.
\end{split}
\label{eq:variationActionTetradGL}
\end{equation}
Here $T^{\mu}_{\;\;a}$ must be the symmetric energy-momentum tensor as follows from comparison with \eqref{eq:variationActionTetrad} for $\delta \hat g_{ab}=0$. Moreover,  when keeping the frame field fixed, $\delta e_{\mu}^{\; a}=0$, we can compare to  \eqref{eq:EMTensorfrommetricVariation} and find that
\begin{equation}
 \hat T^{ab} = \left[T^{\mu\nu} - (1/d) g^{\mu\nu} T^\rho_{\phantom{\rho}\rho}\right] e_\mu^{\;\;a} e_\nu^{\;\;b},
\end{equation}
must be the trace-free part of the symmetric energy-momentum tensor in the unimodular frame.

By using \eqref{eq:variationGLConnection} and eq.\ \eqref{eq:variationConnection} together with the variation  in \eqref{eq:variationActionTetradGeneralizedSpinConnection} and comparing to \eqref{eq:variationActionTetradGL} we find the following relation
\begin{equation}
\begin{split}
T^{\mu\nu} = & \mathscr{T}^{\mu\nu} + \frac{1}{4}\nabla_\rho \left[ 
   \mathscr{S}^{\rho\mu\nu}
+ \mathscr{S}^{\mu\nu\rho} \right.\\
& \left.- \mathscr{S}^{\mu\rho\nu}
-  \mathscr{S}^{\rho\nu\mu} 
+ \mathscr{S}^{\nu\mu\rho}
-  \mathscr{S}^{\nu\rho\mu}
 \right] \\
 & - \frac{1}{2} \left[ N_{\rho\phantom{\rho}\sigma}^{\phantom{\rho}\rho} \mathscr{S}^{\sigma\nu\mu} - N_{\rho\sigma}^{\phantom{\rho\sigma}\nu} \mathscr{S}^{\rho\sigma\mu} + N_{\rho\phantom{\mu}\sigma}^{\phantom{\rho}\mu} \mathscr{S}^{\rho\nu\sigma} \right] \\
 = & \mathscr{T}^{\mu\nu} + \frac{1}{2} \nabla_\rho \left[ S^{\rho\mu\nu} + S^{\mu\nu\rho} + S^{\nu\mu\rho} \right] \\
 & - \frac{1}{2} \left[ N_{\rho\phantom{\rho}\sigma}^{\phantom{\rho}\rho} \mathscr{S}^{\sigma\nu\mu} - N_{\rho\sigma}^{\phantom{\rho\sigma}\nu} \mathscr{S}^{\rho\sigma\mu} + N_{\rho\phantom{\mu}\sigma}^{\phantom{\rho}\mu} \mathscr{S}^{\rho\nu\sigma} \right] .
\end{split}
\label{eq:TUContributions1}
\end{equation} 
In the second equation we have used eq.\ \eqref{eq:SpinCurrent} and $S^{\rho\mu\nu}$ is the spin current. We recognize the Belinfante-Rosenfeld relation between symmetric and canonical energy-momentum tensor as seen before in eq.\ \eqref{eq:EnergyMomentumBelinfanteRosenfeld}, corrected by terms proportional to the distortion tensor. The anti-symmetric part of \eqref{eq:TUContributions1} gives eq.\ \eqref{eq:divergenceSpinCurrent}.

Similarly we find using \eqref{eq:QWInTermsOfHyM}
\begin{equation}
\begin{split} 
 \hat T^{\mu\nu}(x) = & \hat{\mathscr{U}}^{\mu\nu}(x) + \frac{1}{2} \nabla_\rho Q^{\rho\mu\nu}.
\end{split}
\label{eq:TUContributions2}
\end{equation}
This is actually eq.\ \eqref{eq:divergenceQ} obtained previously.

\paragraph*{Local GL$(d)$  transformations.} In a next step let us consider local GL$(d)$  transformations. We use
\begin{equation}
\begin{split}
    \delta e_\mu^{\phantom{\mu}a} = & \left[\delta \omega^a_{\phantom{\mu}b}+\delta \zeta^a_{\phantom{\mu}b} + \delta \zeta \delta^a_{\phantom{a}b} \right] e_\mu^{\phantom{\mu}b}, \\
    \delta \Omega_{\mu\phantom{a}b}^{\phantom{\mu}a} = & - \mathcal{D}_\mu \left[\delta \omega^a_{\phantom{\mu}b}+\delta \zeta^a_{\phantom{\mu}b} + \delta \zeta \delta^a_{\phantom{a}b}\right], \\
    \delta \hat g_{ab} = & - \left[ \delta \omega_{ab} + \delta\zeta_{ab} + \delta\omega_{ba} + \delta\zeta_{ba} \right],
\end{split}
\end{equation}
in eq.\ \eqref{eq:variationActionTetradGL} and find after partial integration
\begin{equation}
\begin{split}
 \delta \Gamma = \int d^d x \sqrt{g} {\bigg \{} &
\left[\mathscr{T}^b_{\;\phantom{b}a} - \frac{1}{2} \mathcal{D}_\mu \mathscr{S}^{\mu\phantom{a}b}_{\phantom{\mu}a}- \hat{\mathscr{U}}_a^{\;b}\right] \\
& \times
\left[\delta \omega^a_{\phantom{\mu}b}+\delta \zeta^a_{\phantom{\mu}b} + \delta \zeta \delta^a_{\phantom{a}b} \right]{\bigg \}}.
\end{split}
\end{equation}
This variation must vanish, so that we find the conservation type relation
\begin{equation}
\mathcal{D}_\mu \mathscr{S}^{\mu\phantom{a}b}_{\phantom{\mu}a} = 2 \left[ \mathscr{T}^b_{\;\phantom{b}a} - \hat{\mathscr{U}}_a^{\phantom{a}b} \right].
\label{eq:conservationTypeRelGLd}
\end{equation}
This should be understood as the Noether relation for GL$(d)$ seen as an extended symmetry. 

We note that eq.\ \eqref{eq:conservationTypeRelGLd} is fully covariant, both in the sense of general coordinate changes, and in the sense of general linear changes of frame in the tangent space. One may decmpose \eqref{eq:cocovariantDerivunimodularmetric} into separate relations for the spin current, shear current and Weyl current, but must be careful to take eq.\ \eqref{eq:cocovariantDerivunimodularmetric} into account.

\paragraph*{General coordinate transformations.}
It is also instructive to study general coordinate transformations (diffeomorphisms) directly in the general linear frame. To that end we start again from eq.\ \eqref{eq:variationActionTetradGL} and use
\begin{equation}
\begin{split}
    & \delta e_\mu^{\phantom{\mu}a} =  \varepsilon^\rho \nabla_\rho e_\mu^{\;\;a} + (\nabla_\mu \varepsilon^\rho) e_\rho^{\;\;a} = - \varepsilon^\rho \Omega_{\rho\phantom{a}b}^{\phantom{\rho}a} e_\mu^{\;\;b} + (\nabla_\mu \varepsilon^\rho) e_\rho^{\;\;a}, \\
    & \delta \Omega_{\mu\phantom{a}b}^{\phantom{\mu}a}  = \varepsilon^\rho \nabla_\rho \Omega_{\mu\phantom{a}b}^{\phantom{\mu}a} + (\nabla_\mu \varepsilon^\rho) \Omega_{\rho\phantom{a}b}^{\phantom{\mu}a},\\
    & \delta \hat g_{ab} = \varepsilon^\rho \partial_\rho \hat g_{ab} = \varepsilon^\rho (\Omega_{\rho ab} + \Omega_{\rho ba} - 2 B_{\rho ab}).
\end{split}
\end{equation}
The change in the action becomes accordingly
\begin{equation}
\begin{split}
\delta \Gamma = \int d^d x \sqrt{g} \varepsilon^\rho {\bigg \{ } & - \nabla_\mu \mathscr{T}^\mu_{\phantom{\mu}\;\rho} - \hat B_{\rho \phantom{a}b}^{\phantom{\mu}a} \hat{\mathscr{U}}_a^{\;b}\\
& - \Omega_{\rho\phantom{a}b}^{\phantom{\rho}a} \left[\mathscr{T}^b_{\phantom{b}\;a} - \frac{1}{2} \nabla_\mu \mathscr{S}^{\mu\phantom{a}b}_{\phantom{\mu}a}- \hat{\mathscr{U}}_a^{\;b} \right] \\
& - \frac{1}{2} \mathscr{S}^{\mu\phantom{a}b}_{\phantom{\mu}a} \left[ \nabla_\rho \Omega_{\mu\phantom{a}b}^{\phantom{\mu}a} - \nabla_\mu \Omega_{\rho\phantom{a}b}^{\phantom{\mu}a} \right]
{\bigg \}}.
\end{split}\label{eq:variationGammadiffeoGLd}
\end{equation}
Using here \eqref{eq:conservationTypeRelGLd} with 
\begin{equation}
\begin{split}
\mathcal{D}_\mu \mathscr{S}^{\mu\phantom{a}b}_{\phantom{\mu}a} = & \nabla_\mu \mathscr{S}^{\mu\phantom{a}b}_{\phantom{\mu}a} - \Omega_{\mu\phantom{c}a}^{\phantom{\mu}c} \mathscr{S}^{\mu\phantom{c}b}_{\phantom{\mu}c} + \Omega_{\mu\phantom{c}c}^{\phantom{\mu}b} \mathscr{S}^{\mu\phantom{a}c}_{\phantom{\mu}a} \\ & - d B_\mu \mathscr{S}^{\mu\phantom{a}b}_{\phantom{\mu}a}, 
\end{split}
\end{equation}
allows to derive from the fact that the variation in \eqref{eq:variationGammadiffeoGLd} must vanish the modified conservation law for the canonical energy-momentum tensor
\begin{equation}
\begin{split}
& \nabla_\mu \mathscr{T}^\mu_{\;\;\;\rho} + \hat B_{\rho\phantom{a}b}^{\phantom{\rho}a} \hat{\mathscr{U}}_a^{\;b} + \frac{d}{2} \Omega_{\rho\phantom{a}b}^{\phantom{\rho}a} B_\mu \mathscr{S}^{\mu\phantom{a}b}_{\phantom{\mu}\;a}  = \frac{1}{2} \mathscr{S}^{\mu\phantom{a}b}_{\phantom{\mu}\;a} \overline{R}^a_{\phantom{a}b\mu\rho}.
\end{split}\label{eq:EMConservationUnimodular}
\end{equation}
On the right hand side we are using the curvature full curvature tensor in the unimodular frame as defined in eq.\ \eqref{eq:CartanRiemann}.
At it should be, eq.\ \eqref{eq:EMConservationUnimodular} reduced to a real conservation law in flat space, and for vanishing non-metricity and contorsion.

In summary, in the unimodular frame one can see nicely the full transformation   group of the frame bundle GL$(d)$. The associated (non-conserved) Noether current is the hypermomentum current with the corresponding divergence-type requation of motion given in eq.\ \eqref{eq:conservationTypeRelGLd}.

\section{Example: scalar field theory with non-minimal coupling}
\label{sec:NonMinimallyCoupledScalarField}

In this section we will discuss an example for an effective action and the resulting construction of the different tensor fields defined in section \ref{sec:GeneralConnection}. The example is illustrative and rather simple. We take the effective action for a single real scalar field to be
\begin{equation}
    \Gamma= \int d^d x  \sqrt{g}   \left\{ -\frac{1}{2}  g^{\mu\nu} \partial_\mu \varphi  \partial_\nu \varphi    - U(\varphi) - \frac{1}{2} \xi R \varphi^2  \right\}.
\label{eq:actionScalarField1}
\end{equation}
Here, $U(\varphi)$ is the effective potential, $R = g^{\sigma\nu}R_{\sigma\nu} =g^{\sigma\nu} R^\rho_{\phantom{\rho}\sigma\rho\nu}$ is the Ricci scalar and $\xi$ denotes its non-minimal coupling to the scalar field $\varphi$. For clarity let us note that the Einstein-Hilbert action would be in our conventions $S_\text{EH}=\int d^d x \sqrt{g}R/(16\pi G_\text{N})$. From dimensional analysis the scaling dimension of $\varphi$ follows as $\Delta_\varphi = (d-2)/2$.

The non-minimal coupling has the value $\xi=(d-2)/(4d-4)$ when the action \eqref{eq:actionScalarField1} has a conformal symmetry. This value can also be seen as a renormalization group fixed point. On the other side, $\xi=0$ is not a renormalization group fixed point and thus $\xi$ is generated by quantum fluctuations even if it should be absent in the microscopic action.

Let us note that as an effective action \eqref{eq:actionScalarField1} should be seen as an approximation. In particular one can expect that quantum fluctuation induce more complex kinetic terms, higher order derivatives, more involved couplings to the curvature tensor, as well as non-local terms. Nevertheless, we can use the model in \eqref{eq:actionScalarField1} for some illustrations.

Let us first consider \eqref{eq:actionScalarField1} in the context of strictly Riemannian geometry and determine the energy-momentum tensor according to eq.\ \eqref{eq:EMTensorfrommetricVariation}. This leads to the so-called improved energy-momentum tensor \cite{Callan:1970ze}, see also \cite{Forger:2003ut}, 
\begin{equation}
\begin{split}
    T^{\mu\nu} = & \partial^{\mu} \varphi \partial^{\nu} \varphi - g^{\mu\nu} \left(  \frac{1}{2} g^{\rho\sigma} \partial_\rho \varphi   \partial_\sigma \varphi    + U(\varphi) + \frac{1}{2}\xi R \varphi^2\right) \\ &
    +\xi \left( R^{\mu\nu} \varphi^2  + g^{\mu\nu} \nabla^\rho \nabla_\rho \varphi^2 - \nabla^{\mu}\nabla^{\nu} \varphi^2 \right).
\end{split}
\end{equation}

Let us now extend eq.\ \eqref{eq:actionScalarField1} to a geometry with general affine connection as discussed in section \ref{sec:GeneralConnection}. This amounts to taking the connection as independent of the metric or alternatively to introduce contorsion, the Weyl gauge field and proper non-metricity. The action \eqref{eq:actionScalarField1} becomes\footnote{This assumes that only quantities with a direct geometric significance like the curvature tensor and its contractions can appear in the effective action. More possibilities arise when torsion and non-metricity can couple to matter seperately, and further investigations are needed to settle when that is the case.}
\begin{equation}
\Gamma = \int d^d x  \sqrt{g}   \left\{ - \frac{1}{2}   g^{\mu\nu} \overline\nabla_\mu \varphi  \overline\nabla_\nu \varphi    - U(\varphi) - \frac{1}{2} \xi \overline R \varphi^2\right\},
\end{equation}
with the co-covariant derivative acting on a scalar field like 
\begin{equation}
\overline\nabla_\mu \varphi = (\partial_\mu  - \Delta_\varphi B_{\mu} )\varphi = \left( \partial_\mu - \frac{d-2}{2} B_\mu \right) \varphi.
\end{equation}
The Ricci scalar $\overline{R}$ is given in eq.\ \eqref{eq:RicciScalarAffine} and its variation in \eqref{eq:variationRicciScalarAffine}. We recall also the connection between the Weyl gauge field and the connection in \eqref{eq:WeylGaugeFieldInTermsOfConnection}.  Similarly one finds from \eqref{eq:variationConnection2}
\begin{equation}
\delta B_{\mu} =  \frac{1}{d} \left[ \delta\Gamma\indices{_\mu^\rho_\rho}- \frac{1}{2} g^{\rho\sigma} \nabla_{\mu} \delta g_{\rho \sigma} \right].
\end{equation}

The (non-conserved) tensor $ \mathscr{U}^{\mu\nu}$ and the hypermomentum tensor $\mathscr{S}^{\mu\phantom{\rho}\sigma}_{\phantom{\mu}\rho}$ are defined through eq.\ \eqref{eq:variationActionMetricConnection}. 
The variation of the action with respect to the metric at fixed connection yields (evaluated at vanishing contorsion and non-metricity)
\begin{equation}
\begin{split}
\mathscr{U}^{\mu\nu} = & 
\partial^{\mu} \varphi \partial^{\nu} \varphi - g^{\mu\nu} \left(  \frac{1}{2} g^{\rho\sigma} \partial_\rho \varphi \partial_\sigma \varphi    + U(\varphi) + \frac{1}{2}\xi R \varphi^2\right) \\& 
    +\xi R^{\mu\nu} \varphi^2 
    + \frac{d-2}{2d} g^{\mu\nu} \nabla_\rho ( \varphi \partial^\rho \varphi).
\end{split}
\end{equation}
Similarly the variation with respect to the connection at fixed metric yields the hypermomentum current,
\begin{equation}
\mathscr{S}\indices{^\mu_\rho^\sigma}= -\frac{d-2}{2d} \delta^\sigma_{\phantom{\sigma}\rho} \, \partial^{\mu} \varphi^2
-\xi g^{\mu\sigma} \partial_{\rho} \varphi^2  +\xi \delta_{\phantom{\mu}\rho}^{\mu} \partial^\sigma \varphi^2.
\end{equation}
The spin tensor can be obtained from this trough eq.\ \eqref{eq:SpinCurrent},
\begin{equation}
S^{\mu\rho\sigma} = \frac12 (\mathscr{S}\indices{^\mu^\rho^\sigma}-\mathscr{S}\indices{^\mu^\sigma^\rho}  )
=-\xi g^{\mu\sigma} \partial^{\rho} \varphi^2  +\xi g^{\mu\rho} \partial^\sigma \varphi^2 .
\end{equation}

The shear and dilation currents follow through eq.\ \eqref{eq:QWInTermsOfHyM}. We find for the dilatation or Weyl current
\begin{equation}
W^\mu = \left(\xi \frac{2d-2}{d} -\frac{d-2}{2d} \right) \partial^{\mu} \varphi^2,
\end{equation}
and for the shear current
\begin{equation}
Q\indices{^\mu^\rho^\sigma} = -\xi g^{\mu \sigma} \partial^\rho \varphi^2
    - \xi g^{\mu\rho} \partial^{\sigma} \varphi^2 
 + \xi  \frac{2}{d} g^{\rho\sigma} \partial^\mu \varphi^2.
\end{equation}
We note that the dilatation current vanishes for the conformal choice $\xi=(d-2)/(4d-4)$. We also note that the shear tensor is in general non-zero. However, it is proportional to gradients and should therefore vanish in many equilibrium situations.

\section{Conclusions}
\label{sec:Conclusions}

We have developed here a formalism to determine expectation values as well as correlation functions of Noether currents from the quantum effective action. These contain immediately all contributions from quantum fluctuations. Technically, the method works with external gauge fields on which the quantum effective action depends in addition to expectation values of matter fields.

The method is very versatile and can be used in particular for the standard, conserved Noether currents associated to global symmetries of the quantum effective action. However, it can also be used for a new class of transformations that have been called ``extended symmetries'' \cite{Canet:2014cta, Tarpin:2017uzn, Tarpin:2018yvs}, under which the action is not invariant but changes by a term that is locally known. More precisely, this change must be proportional to a term that is actually known at the macroscopic level of the quantum effective action in order for the transformation to be useful. Associated to such ``extended symmetries'' one finds non-conserved ``Noether currents''. Their equation of motion has a form similar to a covariant conservation law but with a non-vanishing and known term on the right hand side. Note that the situation is very similar to an anomalous symmetry, where the violation of the conservation law is proportional to the anomaly. 

After a general discussion of this construction we turned to applications of these ideas in the context of space-time geometry. First, the symmetry under general coordinate transformations leads as usual to the covariant conservation of the symmetric energy-momentum tensor. More interesting are further transformations corresponding to changes of basis in the frame- and spin bundle. In particular we recalled the treatment of local internal Lorentz transformations (including rotations and boosts), local dilatations or Weyl transformations, and also discussed the less-known local space-time shear transformations. The associated currents are the spin current, the dilatation or Weyl current and the shear current. Together they form a rank-three tensor known as hypermomentum current \cite{1976ZNatA..31..111H,1976ZNatA..31..524H,PhysRevD.17.428,Hehl:1994ue}. The latter can also be understood as the (non-conserved) Noether current associated to GL$(d)$ transformations in the frame bundle. In particular for the shear current we derived a new divergence-type equation of motion.

It is only under special circumstances that real conservation laws arise. For example, the Weyl current is conserved in the presence of a conformal symmetry (but then typically vanishes). Or the spin current is conserved when the canonical energy-momentum tensor is symmetric. (Our formalism yields an expression for the canonical energy-momentum tensor in terms of a variation of the effective action.) The shear current is usually not conserved, except when it vanishes.

An interesting application of the insights gained here might concern relativistic fluid dynamics. While the usual formulation builds up on the covariant conservation law for the energy-momentum tensor, additional equation of motion are available in our formalism, and their connection to the quantum effective action is now understood. It is very interesting that for a given quantum effective action the currents themselves are known, as well as their correlation functions, at least in principle. On the other side, the state dependence of the quantum effective action might be carried to rather good approximation by the (fluid dynamic) degrees of freedom of the energy-momentum tensor, by the matter field expectation values, and additionally by the components of the hypermomentum current. This could lead to a rather powerful formalism for quantum field dynamics out-of-equilibrium. Understanding how the components of the hypermomentum tensor evolve might be of interest for many situations in non-equilibrium quantum field theory, such as in condensed matter physics, heavy ion collisions, or cosmology.

We believe that in particular the dilatation current and the shear current are interesting because their divergence-type equation of motion could give the evolution equations for the non-equilibrium degrees of freedom related to bulk and shear viscous dissipation. Moreover, the structure of the equations of motion is such that these equations could actually be causal in the relativistic sense as explained in  ref.\ \cite{Geroch:1990bw}. In this regard, our equations are similar to the equations of motion for the so-called divergence-type theories of relativistic fluid dynamics. We plan to investigate these matters in more detail in a forthcoming publication.

Technically we obtain equations of motion for the components of the hypermomentum tensor by varying the affine connection independent of the metric, tetrad or frame field. Conceptually this amounts to varying the non-Riemannian parts of space-time geometry, specifically contorsion, the Weyl gauge field and proper non-metricity. It is important to note here that for us the non-Riemannian geometry is a purely calculational device. After the variations are done we can evaluate all expressions at vanishing contorsion and non-metricity, i.\ e.\ in the (pseudo) Riemannian geometry of general relativity. However, from the point of view of theories of modified gravity (beyond Einsteins theory of general relativity), our findings may also be of interest. Specifically, it has been argued that the spin current, dilatation current and shear current are natural source terms to appear in such extended theories of gravity \cite{Kibble:1961ba,Sciama:1964wt,1976ZNatA..31..111H, 1976ZNatA..31..524H, Hehl:1976kv, PhysRevD.17.428, Hehl:1994ue, Capozziello:1996bi, Puetzfeld:2004yg, Vitagliano:2010sr, Blagojevic:2012bc,BeltranJimenez:2018vdo,Shimada:2018lnm, Jimenez:2019ghw, Iosifidis:2020gth}. It is therefore useful to understand well under which circumstances they are non-zero.

In the present paper we have focused entirely on (quantum) field theory. However, in light of our findings it might also be interesting to revisit actions for particles (or strings or branes), and to investigate wether and how contributions to the shear current, Weyl current and spin current would arise from variations in geometry there.

On the example of a scalar field theory with non-minimal coupling to gravity we have shown that all components of the hypermomentum tensor can be non-vanishing. However, they are proportional to gradients and might therefore vanish in many equilibrium situations. We believe that quantum fluctuations indeed induce typically a non-vanishing shear and dilatation current in non-equilibrium situations. 

The influence of quantum and statistical fluctuations on the non-equilibrium currents can be investigated further, for example with the functional renormalization group \cite{Wilson:1971bg, Wegner:1972ih, Polchinski:1983gv, Wetterich:1992yh}, and we plan to do so. 

\section*{Acknowledgments}
This work is supported by the Deutsche Forschungsgemeinschaft (DFG, German Research Foundation) under Germany's Excellence Strategy EXC 2181/1 - 390900948 (the Heidelberg STRUCTURES Excellence Cluster), SFB 1225 (ISOQUANT) as well as FL 736/3-1, and by U.S. Department of Energy, Office of Science, Office of Nuclear Physics, grants Nos.
DE-FG-02-08ER41450.

\begin{appendix}
\section{Lie algebra of GL(d)}\label{sec:appLieAlgebra}
The Lie algebra of the general linear group GL(d) can be decomposed into different sectors. In the fundamental representation one may work with $(M^{AB})^C_{\phantom{C}D} = -i (\eta^{AC} \delta^B_{\;\;D} - \eta^{BC} \delta^A_{\;\;D})$ for the generators of Lorentz transformations, $D^A_{\phantom{A}B}=-i \delta^A_{\phantom{A}B}$ for the generator of dilatations, and $(S^{AB})^C_{\phantom{C}D} = -i (\eta^{AC} \delta^B_{\;\;D} + \eta^{BC} \delta^A_{\;\;D}-(2/d)\eta^{AB}\delta^C_{\;\;D})$ for the generators of shear transformations.

The Lie brackets are then
\begin{align}
    \left[ M^{AB}, M^{CD}\right] = & -i {\big (} \eta^{BC} M^{AD} - \eta^{AC} M^{BD} \nonumber \\& + \eta^{BD} M^{CA} - \eta^{AD} M^{CB} {\big )}, \\
\left[ M^{AB}, S^{CD} \right] = & -i {\big (} \eta^{BC} S^{AD} - \eta^{AC} S^{BD} \nonumber \\& + \eta^{BD} S^{CA} - \eta^{AD} S^{CB} {\big )}, \\
\left[ S^{AB}, S^{CD}\right] = & -i {\big (} \eta^{BC} M^{AD} + \eta^{AC} M^{BD} \nonumber\\& - \eta^{BD} M^{CA} - \eta^{AD} M^{CB} {\big )},\\
\left[ M^{AB}, D\right] = & \left[ S^{AB}, D\right] = [D,D]=0.
\label{eq:LieBracketsGL}
\end{align}
One observes that Lorentz transformations and dilations form subgroups, while shear transformations do not.

\end{appendix}

\end{document}